\documentclass[useAMS,usenatbib,usegraphicx]{mn2e}

\usepackage{times}

\newcommand{\JHK}{JHK_{\rm s}}
\newcommand{\Ks}{K_{\mathrm{s}}}
\newcommand{\AK}{A_{K_{\mathrm{s}}}}
\newcommand{\MK}{M_{K_{\mathrm{s}}}}
\newcommand{\DMAK}{{(\mu_0,~\AK)}}
\newcommand{\Nobs}{N_{\mathrm{obs}}}
\newcommand{\Ndet}{N_{\mathrm{det}}}
\newcommand{\PG}{P_{\mathrm{G01}}}
\newcommand{\PI}{P_{\mathrm{IRSF}}}
\newcommand{\MuLMC}{\mu_{\mathrm{LMC}}}
\newcommand{\Pf}{{\rm Pflag}}
\newcommand{\Mf}{{\rm Mflag}}
\newcommand{\IRACa}{{[3.6]}}
\newcommand{\IRACb}{{[4.5]}}
\newcommand{\IRACc}{{[5.8]}}
\newcommand{\IRACd}{{[8.0]}}

\title[Miras towards the Galactic centre]
{A near-infrared survey of Miras and the distance to the Galactic Centre}
\author[N. Matsunaga~{et~al.}]
{Noriyuki Matsunaga$^{1,2}$\thanks{matsunaga@ioa.s.u-tokyo.ac.jp}\thanks{Research Fellow of the Japan Society for the Promotion of Science},
Takahiro Kawadu$^{2}$,
Shogo Nishiyama$^{2}$\footnotemark[2],
\newauthor
Takahiro Nagayama$^{3,2}$,
Hirofumi Hatano$^{3}\footnotemark[2]$,
Motohide Tamura$^{4}$,
\newauthor
Ian S. Glass$^{5}$
and Tetsuya Nagata$^{2}$
\\
$^{1}$ Institute of Astronomy, University of Tokyo, 2-21-1 Osawa, Mitaka, Tokyo 181-0015, Japan \\
$^{2}$ Department of Astronomy, Kyoto University, Kitashirakawa-Oiwake-cho, Sakyo-ku, Kyoto 606-8502, Japan \\
$^{3}$ Department of Astrophysics, Nagoya University, Furo-cho, Chikusa-ku, Nagoya 464-8602, Japan \\
$^{4}$ National Astronomical Observatory of Japan, 2-21-1 Osawa, Mitaka, Tokyo 181-8588, Japan \\
$^{5}$ South African Astronomical Observatory, PO Box 9, Observatory 7935,
South Africa
}

\begin{document}

\date{Accepted -. Received -; in original form 2009 May 14}

\pagerange{\pageref{firstpage}--\pageref{lastpage}} \pubyear{2009}

\maketitle

\label{firstpage}

\begin{abstract}
We report the results of a near-infrared survey for long-period variables in a
field of view of 20~arcmin by 30~arcmin towards the Galactic Centre (GC).
We have detected 1364 variables, of which 348 are identified with those
reported in \citet{Glass-2001}. We present a catalogue and photometric
measurements for the detected variables and discuss their nature. We also
establish a method for the simultaneous estimation of distances and
extinctions using the period-luminosity relations for the $\JHK$ bands. Our
method is applicable to Miras with periods in the range 100 -- 350~d and
mean magnitudes available in two or more filter bands. While $J$-band means
are often unavailable for our objects because of the large extinction, we
estimated distances and extinctions for 143
Miras whose $H$- and $\Ks$-band
mean magnitudes are obtained. We find that most are located at the same
distance to within our accuracy. Assuming that the barycentre of these Miras
corresponds to the GC, we estimate its distance modulus to be
$14.58 \pm 0.02~({\it stat.}) \pm 0.11~({\it syst.})$~mag, corresponding to
$8.24\pm 0.08~({\it stat.}) \pm 0.42~({\it syst.})$~kpc. We have assumed the
distance modulus to the LMC to be 18.45~mag, and the uncertainty in this
quantity is included in the systematic error above. We also discuss the
large and highly variable extinction. Its value
ranges from 1.5~mag to larger than 4~mag in $\AK$ except towards the thicker
dark nebulae and it varies in a complicated way with the line of sight. We
have identified mid-infrared counterparts in the {\it Spitzer}/IRAC
catalogue of \citet{Ramirez-2008} for most of our variables and find that
they follow rather narrow period-luminosity relations in the 3.6 to
8.0~$\mu\mathrm{m}$ wavelength range.
\end{abstract}


\section{Introduction\label{sec:Intro}}

The detailed structure of the Galaxy remains to be revealed. It is difficult
to get a clear picture of its shape because the Sun and the Earth themselves
are located within the Galactic disc. Furthermore, strong interstellar
extinction towards the Galactic disc hinders
our understanding of Galactic
structure. Several kinds of tracers have been used to investigate this
structure using various observational methods: interstellar gas (Nakanishi
and Sofue, \citeyear{Nakanishi-2003}, \citeyear{Nakanishi-2006}) and
red-clump stars (\citealt{Nishiyama-2005};
\citealt{Babusiaux-2005}; \citealt{Rattenbury-2007}),
for example. The tracers we make use
of in the present paper are Miras, i.e. long-period variables with large
amplitudes ($V > 2.5$~mag) and relatively regular light curves ($P \geq
100$~d).

The period-luminosity relation (PLR) of the Miras has been widely used as a
distance indicator since \citet{Glass-1981} and \citet{Feast-1989}
established it for the Miras in the Large Magellanic Cloud (LMC). Our
South African -- Japanese collaboration has recently applied it to
a few dwarf spheroidal galaxies of the Local Group
(\citealt{Menzies-2008}, for Phoenix; \citealt{Whitelock-2009}, for Fornax).
\citet{Rejkuba-2004} has shown that the PLR can be also applied
to the Miras in Cen~A, a peculiar elliptical galaxy beyond the Local Group.

We can also investigate the distribution of Miras in the Galaxy by
making use of the PLR
(\citealt{Groenewegen-2005};
Matsunaga, Fukushi \& Nakada, 2005, here after M05).
An important advantage of them as tracers is that we
can obtain the position of each individual. M05 have
conducted an exploratory study using the OGLE-II variability catalogue and
the 2MASS all-sky catalogue. They presented the distribution of Miras in and
around the bulge, restricted however to the lines of sight of the OGLE-II
survey \citep{Wozniak-2002}. Nevertheless, it was shown that such a map can be 
used to study Galactic structure directly.  Based on this method, kinematic
information can easily be combined with positional information and it is
possible to locate each tracer in 6-dimensional phase space. This is
different from the case for red-clump stars, for instance, because their
distribution needs to be solved for as a group in a statistical way.
Although a survey of Miras requires a long-term monitoring programme, we
have found that it is a powerful tool for revealing Galactic structure.

In this paper we present the results of our near-infrared (near-IR) survey
of Miras towards the Galactic Centre (hereafter GC).
\citeauthor{Glass-2001}~(2001, hereafter G01) conducted
a $K$-band~(2.2~$\mu$m) survey over almost the same region as ours and
found
409 long-period variables. Our survey is deeper by more than one
magnitude and, more importantly, includes simultaneous monitoring in
$J~(1.25~\mu\mathrm{m}$) and $H~(1.63~\mu\mathrm{m}$)
as well as $\Ks~(2.14~\mu\mathrm{m})$.
We have detected 1364 long-period variables and compiled their catalogue
including all the available photometric measurements.
From the catalogue we select Miras and investigate Galactic structure.
Our first step has been to establish the validity of the Mira PLR method for
estimating both extinction and distance. In previous studies, e.g.\  M05,
researchers adopted interstellar extinction values
found by other methods in order to estimate distances of Miras based on the
$\log~P$-$K$ (or -$m_{\mathrm{bol}}$) relation. However, the extinction
varies in a complicated way across the Galactic plane and is difficult to
estimate in many cases. Both extinction and distance can be obtained
simultaneously by using the PLR in two or more near-IR filters as is
presented below. Our second step makes use of the survey data to explore the
distribution of Miras and to estimate the distance to the GC, taken as the
barycentre of the stellar population. We will also discuss interstellar
extinction towards the GC region.

\section{Observation and data reduction}

\subsection{Observations}

We used the IRSF 1.4-m telescope and the SIRIUS near-IR camera for our
monitoring survey. These were established by Nagoya University and National
Astronomical Observatory of Japan and are sited at the Sutherland station of
the South African Astronomical Observatory. They can take images in $\JHK$
simultaneously with a field of view of $7.7 \times 7.7 \; {\rm arcmin}^2$
(for details, see \citealt{Nagashima-1999} and \citealt{Nagayama-2003}).
The seeing size is typically 1.3~arcsec and is around 1~arcsec at its best.

\begin{figure*}
\begin{minipage}{150mm}
\begin{center}
\includegraphics[clip,width=0.85\hsize]{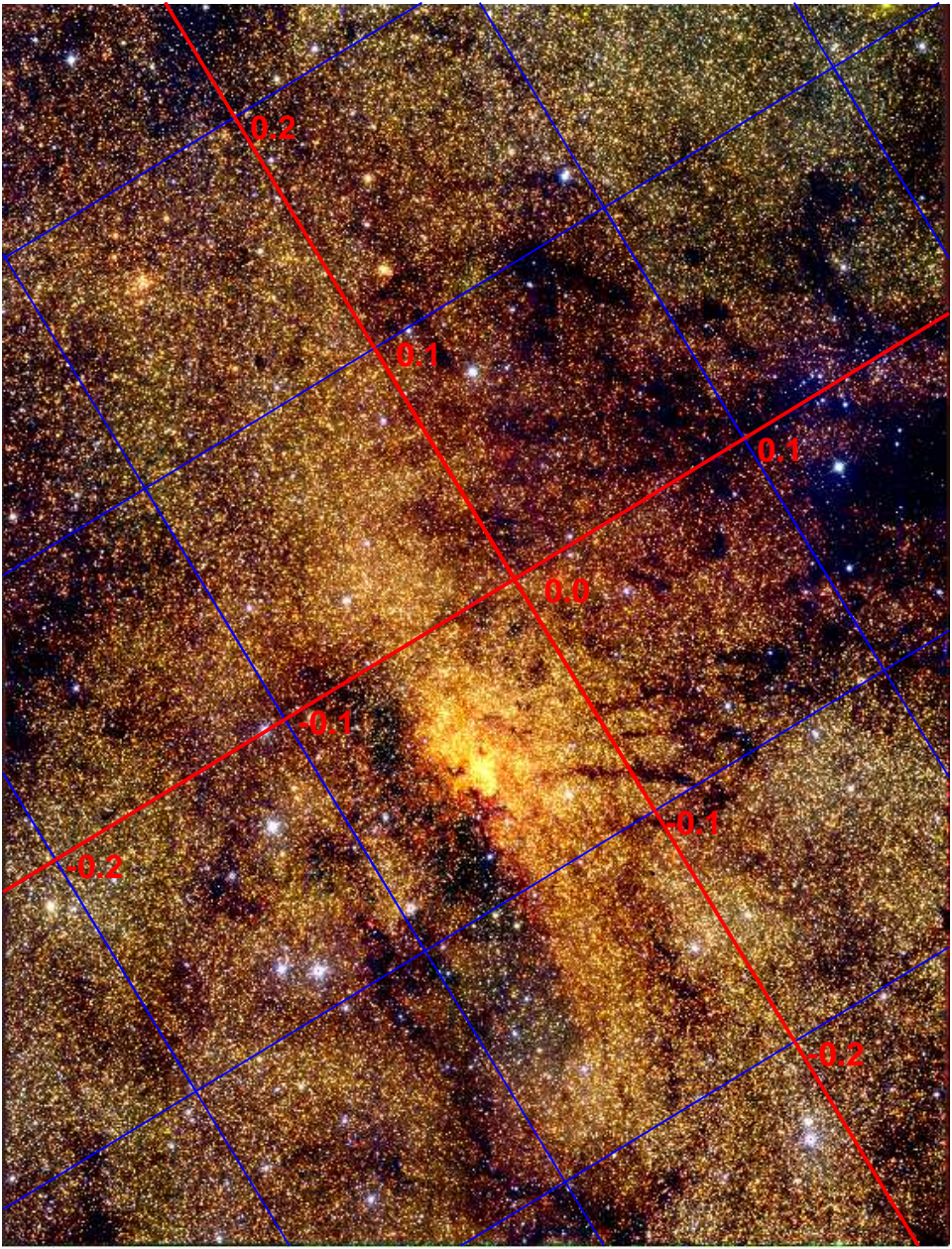}
\end{center}
\caption{
An $H$-band image (or an RGB composite image in the online journal)
of the observed field.
North in the equatorial system is up, and east is to the left.
Solid (or red) lines indicate $l=0^\circ$ and $b=0^\circ$,
while a grid with spacing of $0.1^\circ$ is shown with dashed lines
(or blue ones).
The covered area is about 20~arcmin by 30~arcmin.
\label{fig:chart}}
\end{minipage}
\end{figure*}

We repeatedly observed 12 fields of view around the GC. Their central
coordinates are given in Table~\ref{tab:Fields} together with the numbers of
times each one was observed, $\Nobs$. A large part of the data was obtained
in 2005 and 2006, while additional data were collected between 2001 and
2008. Each set of observations comprises of ten exposures for five seconds
at slightly dithered positions. The number of the observation sets of a
given field obtained per night ranges from one to six. We present an
H-band image (or an RGB-composite image in the online journal)
of the observed region in Fig.~\ref{fig:chart} based on
one set of $\JHK$ images.

\begin{table*}
\begin{minipage}{170mm}
\caption{The observed fields. The central coordinates and the numbers
of observations $\Nobs$ are indicated. We also list the number of sources
($N_{\mathrm{src}}$) and variables ($N_{\mathrm{var}}$) detected in each
field as well as the total number in which we do not include duplicate
detections in neighbouring fields.
\label{tab:Fields}}
\begin{center}
\begin{tabular}{ccccllcccccccccc}
\hline
Field   &  \multicolumn{2}{c}{Centre (J2000.0)} & \multicolumn{2}{c}{Centre (Galactic)} & \multicolumn{8}{c}{$\Nobs$} & $N_{\mathrm{src}}$ & $N_{\mathrm{var}}$ \\
        &  RA        &  Dec   & \multicolumn{1}{c}{$l (^\circ)$} & \multicolumn{1}{c}{$b (^\circ)$}       &   2001 & 2002 & 2004 & 2005 & 2006 & 2007 & 2008 & Total  & & \\
\hline
1745-2900A & 17:46:10.5 & $-$28:53:47.8  & $+0.0970$ & $-0.0831$ 
  &1&1&0&47&42&2&1& 94 & 9,266 & 128\\
1745-2900B & 17:45:40.0 & $-$28:53:47.8  & $+0.0391$ & $+0.0119$ 
  &0&1&0&47&41&2&1& 92 & 6,797 & 132\\
1745-2900C & 17:45:09.5 & $-$28:53:47.8  & $-0.0189$ & $+0.1068$ 
  &1&1&0&45&40&2&1& 90 & 6,041 &  89\\
1745-2900D & 17:46:10.5 & $-$29:00:28.0  & $+0.0020$ & $-0.1409$ 
  &1&1&0&46&42&2&1& 93 & 8,692 & 104\\
1745-2900E & 17:45:40.0 & $-$29:00:28.0  & $-0.0558$ & $-0.0460$ 
  &1&1&0&45&41&2&1& 91 & 8,451 & 225\\
1745-2900F & 17:45:09.5 & $-$29:00:28.0  & $-0.1137$ & $+0.0488$ 
  &1&1&0&43&41&2&1& 89 & 8,185 & 151\\
1745-2900G & 17:46:10.5 & $-$29:07:07.8  & $-0.0928$ & $-0.1987$ 
  &1&1&0&41&41&2&1& 87 & 9,576 & 101\\
1745-2900H & 17:45:40.0 & $-$29:07:07.8  & $-0.1506$ & $-0.1039$ 
  &1&1&0&43&41&2&1& 89 & 7,399 & 123\\
1745-2900I & 17:45:09.5 & $-$29:07:07.8  & $-0.2085$ & $-0.0091$
  &0&1&0&43&36&2&1& 83 & 8,747 & 122\\
1745-2940G & 17:46:10.5 & $-$28:47:07.9  & $+0.1919$ & $-0.0254$
  &1&1&1&41&38&2&1& 85 & 8,885 & 174\\
1745-2940H & 17:45:40.0 & $-$28:47:07.9  & $+0.1339$ & $+0.0698$
  &1&1&1&40&39&2&1& 85 & 9,237 & 130\\
1745-2940I & 17:45:09.5 & $-$28:47:07.9  & $+0.0758$ & $+0.1648$
  &1&1&1&15&40&2&0& 60 & 8,636 &  80\\
\hline
Total      &            &               & & &
  & & &  &  & & &    &81,019 &1,364\\
\hline
\end{tabular}
\end{center}
\end{minipage}
\end{table*}

\subsection{Pipeline reduction\label{sec:Pipeline}}

The raw data were first processed by using the SIRIUS pipeline software
(Yasushi Nakajima, private communication)\footnote{ The software is a set of
scripts making use of {\small IRAF} (Imaging Reduction and Analysis
Facility).
http://www.z.phys.nagoya-u.ac.jp/~nakajima/sirius/software/software.html}.
A rough
outline of the procedures is as follows. First, a dark image was subtracted
from every raw image in order to eliminate the effect of dark current. We
took ten frames with the shutter of the SIRIUS camera closed on every
observing night. These were combined into one dark image. Then we corrected
the pixel-to-pixel variation of sensitivity by dividing dark-subtracted
images by a flat image. The flat image was produced from twilight-flat
exposures. Several pairs of images with different sky levels were collected
from twilight-flat images taken in sequence, and the differences between two
images of the pairs
were combined to produce a flat-field image after
normalization. In order to subtract the background pattern of the images, we
took exposures of a relatively sparse region with central coordinates
17:11:24.7, $-$27:27:20 (J2000.0). We took 10 dithered images of the blank
field and combined them without adding any shifts to compensate for the
dithering offsets. Thus we obtained a sky image without stellar signals but
showing the background pattern of the images; this sky frame we subtracted
from each flat-fielded image. The final step was to combine the 10 dithered
images for a target field into a scientific image. Bright stars were
selected in the sky-subtracted images, and the offsets among the images were
calculated. The average of the images was taken after un-dithering them. The
scientific image thus obtained is then free of spurious noise caused by bad
pixels and cosmic ray events.

\subsection{Photometry\label{sec:Photometry}}

Point-spread-function (PSF) fitting photometry was performed on the
scientific images with the {\small DAOPHOT} package in {\small
IRAF}\footnote{{\small IRAF} is distributed by the National Optical
Astronomy Observatory, which is operated by the Association of Universities
for Research in Astronomy, Inc., under cooperative agreement with the
National Science Foundation.}. For each field one image taken under the best
conditions (of weather and seeing) was selected from among the $\Nobs$
images for each filter as a reference frame.
In order to standardize the magnitudes we compared the photometric result of
the reference frame with the photometric catalogue by \citet{Nishiyama-2006a}.
Their data were taken with the same instrument as ours. We used our own
implementation of the Optimal Pattern Matching (OPM) algorithm
\citep{Tabur-2007} to match stellar positions in our images and celestial
coordinates given by
\citet{Nishiyama-2006a}. The magnitudes we obtained are on the IRSF/SIRIUS
natural system, whose zero-point calibration was based on a standard star in
\citet{Persson-1998}.

\begin{figure}
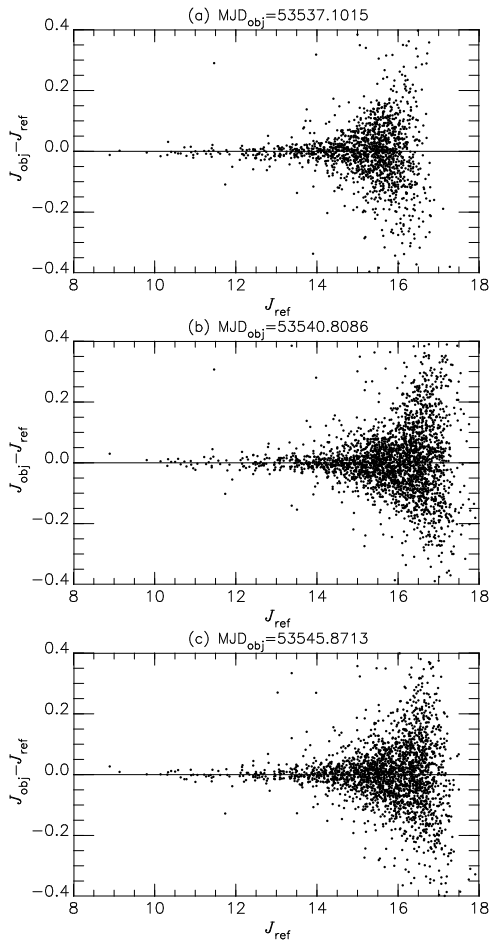

\begin{center}
\begin{minipage}{75mm}
\begin{center}
\includegraphics[clip,width=0.85\hsize]{Fig2a.ps}
\includegraphics[clip,width=0.85\hsize]{Fig2b.ps}
\includegraphics[clip,width=0.85\hsize]{Fig2c.ps}
\end{center}
\caption{
Examples of magnitude comparisons between the reference data
($\mathrm{MJD_{ref}}=53665.7594$ in this case) and repeated observations.
These plots are for the 1745-2900A field and the MJDs of the target data
are indicated on top of the panels.
\label{fig:PhotoComp}}
\end{minipage}
\end{center}
\end{figure}

We also performed PSF-fitting photometry using {\small DAOPHOT}
for the remaining ($\Nobs-1$) images,
and the results were compared with those from the reference frame.
The comparisons were again made using the OPM algorithm.
The positions are consistent among the images to within $\pm 0.1$~arcsec.
We present examples of the magnitude comparisons in
Fig.~\ref{fig:PhotoComp}.
The zero points of the magnitude scale for the ($\Nobs-1$) images
were calibrated to fit with that of the reference
based on 3-$\sigma$ clipped medians.
The differences between magnitudes in the frames
are distributed around zero for vast majority of stars,
whilst the differences for variable stars follow their variations.

\subsection{Variability Detection\label{sec:Detection}}

\begin{figure}
\begin{center}
\begin{minipage}{75mm}
\begin{center}
\includegraphics[clip,width=0.85\hsize]{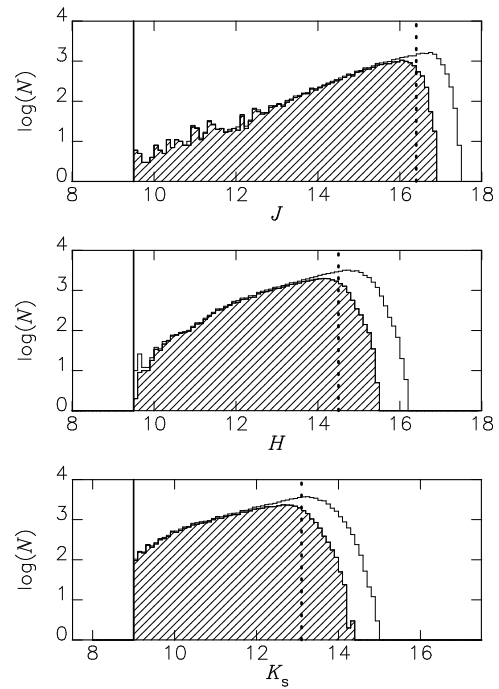}
\end{center}
\caption{
Histogram of $\JHK$ magnitudes of detected sources
including non-variables.
The shaded area is the histogram for sources
with $\Ndet \geq 0.8 \Nobs$ while the outline indicates
that for all the sources included in our source list
($\Ndet \geq 0.5 \Nobs$). The solid lines indicate
the saturation limits while the dotted line do the detection limit
(see text for their definitions).
\label{fig:Mhist}}
\end{minipage}
\end{center}
\end{figure}

Using the procedures described in the preceding section, we obtained lists
of the sources detected in $\Nobs$ images for each field and each filter. We
expect that the faintest objects are detected only in some images due to
variable conditions such as seeing. We included sources which were detected
in $0.5 \Nobs$ or more images in our catalogue. Using $\Ndet$ to
denote the number of detections of a given source, the condition can be
written as $\Ndet \geq 0.5 \Nobs$. The number of detected sources,
$N_{\mathrm{src}}$, is listed in Table~\ref{tab:Fields} for each field. The
total number for the 12 fields is also given. Fig.~\ref{fig:Mhist} shows
histograms of $\JHK$ magnitudes and illustrates the detection limit. The
shaded area is the histogram for sources with $\Ndet \geq 0.8 \Nobs$ while
the outline is that for all the sources ($\Ndet \geq 0.5 \Nobs$). For a
wide range of magnitudes, it is clear that the sources in our catalogue
appear in almost all 
of the $\Nobs$ images. The numbers of
detections start to decrease at the faint end. The detection limit for each
image depends on the seeing and other weather conditions as well as
the crowdedness, which varies considerably as seen in
Fig.\ \ref{fig:chart}. We here define the typical limiting magnitude as that at
which the number of sources with
$\Ndet \geq 0.8 \Nobs$ becomes less than half of those with
$\Ndet \geq 0.5 \Nobs$;
$J=16.4$, $H=14.5$, and $\Ks=13.1$~mag.
For sources fainter than these magnitudes it gets difficult
to keep the detection ratios, $\Ndet/\Nobs$,
larger than 80 percent.
The limits are indicated by the dotted lines in Fig.~\ref{fig:Mhist}

We also found that the brightest sources are affected
by deviations from linear response of the detectors.
The deviation gets larger than 5 percent
for objects brighter than $J=9.0$--$9.5$, $H=9.0$--$9.5$,
or $\Ks=8.5$--$9.0$~mag. This limit is also
dependent on seeing size and background brightness.
In the following we do not consider magnitudes 
brighter than 9.5, 9.5, and 9.0~mag in $J$, $H$, and $\Ks$, respectively.
We did not include measurements above the upper brightness limits in
$\Ndet$. 

\begin{figure}
\begin{center}
\begin{minipage}{75mm}
\begin{center}
\includegraphics[clip,width=0.85\hsize]{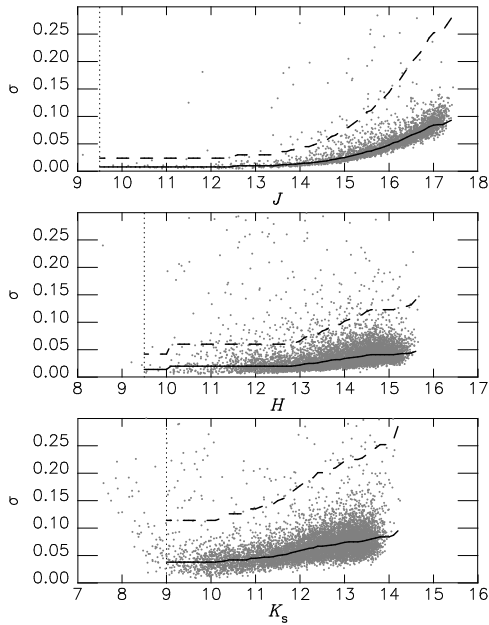}
\end{center}
\caption{
The standard deviation of $\Ndet$ magnitudes is plotted against
the mean magnitude for each object.
The solid curves are the median errors at a given magnitude and
the limits of non-variable stars are taken to be the dashed curves.
The latter are shifted by a factor of three from the solid ones.
See text for details.
\label{fig:MagDev}}
\end{minipage}
\end{center}
\end{figure}

Each source included in our list has magnitudes obtained for $\Ndet$ images
in each band, where $\Ndet$ can be different for each of the $\JHK$ bands.
We calculated the standard deviations (SDs) of the $\Ndet$ magnitudes, which
are plotted against the mean magnitudes in Fig.~\ref{fig:MagDev} for the
case of the 1745-2900A field (as an example). As may be expected, the
fainter sources tend to have the larger SDs. Assuming that the majority of
the sources are not variable, at least to our photometric accuracy, their
SDs arise from photometric errors. We estimated the error as a function of
mean magnitude by taking the median of the SD at each magnitude. The median
was calculated for every 0.1~mag interval, with a width of $\pm 0.2$~mag.
This estimation was carried out from the faint to the bright end and we
imposed the constraint that the error curve is a monotonically increasing
function of magnitude, that is, $\sigma (m_1) < \sigma (m_2)$ if $m_1 <
m_2$ and $\sigma (m_1) = \sigma (m_2)$ otherwise. Examples of the error
curves obtained are indicated by thick solid lines in Fig.~\ref{fig:MagDev}.
The vertical dotted lines indicate the limit brighter than which the
magnitudes are affected by the deviation from linear response.
The measurements over these limits are nevertheless plotted
in Fig.~\ref{fig:MagDev} to
illustrate such an effect. The stars affected have large SDs, 
as seen in the bottom ($\Ks$) panel, and are not considered further.

The SD values above which objects were taken to be 
candidate variables were obtained by
multiplying the error curve by a factor of three and are plotted as the
thick dashed lines in Fig.~\ref{fig:MagDev}. For every candidate, we
inspected the light curve and the appearance of its image by eye in order to
determine whether the variation is real or spurious. The candidates excluded
after inspection often lie near the edges of the images, causing their
photometry to be affected. Some others are severely distorted by
neighbouring bright star(s) that are saturated.  At the same time, we
searched for short-period variables with $0.1 < P < 60$~d among the
candidate variables, which include Cepheids and eclipsing binaries. We found
approximately 50 short-period variables. They are not included in this paper
and will be reported in a forthcoming paper. The number of the remaining
variables detected in each field,
$N_\mathrm{var}$, is given in Table~\ref{tab:Fields}. When variables
were found in two or more neighbouring fields near their edges, we confirmed
that the light curves agreed and adopted only one entry for each variable.
Thus 1364 long-period variables\footnote{We designate variables which are
not of short period ($P<60$) as long-period variables even if their light
curves are not truly periodic. Also see section \ref{sec:amplitude}
regarding the definition of Miras.} were selected for our catalogue.

\section{Catalogue construction\label{sec:Construction}}

In this section we discuss the estimation of periods, mean magnitudes, and
their accuracies. We further carry out cross identifications with the
surveys by G01 and \citet{Ramirez-2008}. The comparison with the G01
catalogue is, in some cases, useful for period estimation.

\subsection{Comparison with \citet{Glass-2001}}

Miras listed in G01, also see \citet{Glass-2002}, were searched for in our
catalogue. In order to adjust any systematic shifts in position between the
two datasets, we added offsets to the positions in the G01 catalogue for the
different G01 fields ($d_{\rm RA}$ and $d_{\rm Dec}$, given in
Table~\ref{tab:correlation}). The application of these offsets led to better
positional agreements. We found 348 matches with a tolerance of 3~arcsec.
The standard deviations of the differences in Right Ascension (RA) and
Declination (Dec) are approximately 0.5~arcsec after the offsets mentioned
above.

G01 listed 12 duplicated detections
of the Miras in their neighbouring fields in their table 3.
In course of the cross identification, we found several more
pairs of objects listed twice by G01.
Their 2-18 and 3-7655 corresponds to 17453194-2857473 in our catalogue,
while 4-6/19-23\footnote{This pair was also identified by \citet{Glass-2002}.}
to 17453003-2905100, 9-75/12-46 to 17461431-2854088,
and 23-1198/24-2 to 17450596-2857483.
We also found that 2-3 is identical to 16-24/17-1.
The counterpart of 16-49 is 3-266,
not 3-226 as listed in their paper.
Additionally both 9-4 and 12-2 are close to one of the IRSF variables,
17461373-2855143,
although they are slightly separated from each other.
Their parameters are close to each other and consistent with
the IRSF counterpart.
We conclude that they are also the same object detected
twice in neighbouring fields.

Among 56 G01 objects whose counterpart was not found, 39 are located outside
our fields of view. Coordinates of the other 17 objects are within our
survey area, but they are not detected as variable stars. We list those not
identified in Table~\ref{tab:unmatchedGlass} with the suspected reasons for
the failure. For example,
the coordinate of 6-112 is rather close to a very bright star
which is saturated in our images, while 6-112 is listed as rather faint. Its
$Q$ flag\footnote{G01 gave a flag between 0 (low) and 3 (high) for each
object to classify the light curves according to their quality.} given in
the G01 catalogue is low and their photometry must have been affected by the
saturated star. Both 12-799 and 12-1236 are separated from 17462073-2853253
by about 3~arcsec. Different parameters are given in G01 for these objects,
and it is not clear which one corresponds to the IRSF variable. Their $Q$
flags are low. We did not consider the counterparts for these two G01
objects.

\begin{table}
\begin{minipage}{70mm}
\begin{center}
\caption{
The 25 fields in the G01 catalogue and the number of variables.
$N_{\mathrm{G01}}$ -- the number of objects reported in G01;
$N_{\mathrm{mat}}$ -- the number of the IRSF ones identified with those in G01;
$d_{\rm RA}$ and $d_{\rm Dec}$
-- offset between the coordinates in G01 and our catalogue.
The total numbers after removing duplicate entries from neighbouring fields
are listed in the last line.
\label{tab:correlation}}
\begin{tabular}{crrrr}
\hline
Field & $N_{\mathrm{G01}}$ & $N_{\mathrm{mat}}$ 
& $d_{\mathrm{RA}}$~(arcsec) & $d_{\mathrm{Dec}}$~(arcsec) \\
 \hline
 GC1 & 14 & 14 & $-0.57$ & $ 1.01$ \\
 GC2 & 26 & 26 & $-0.35$ & $ 1.16$ \\
 GC3 & 35 & 32 & $ 0.39$ & $ 0.50$ \\
 GC4 & 13 & 13 & $ 0.45$ & $ 1.55$ \\
 GC5 & 10 &  7 & $-0.30$ & $ 1.14$ \\
 GC6 & 21 & 20 & $-0.65$ & $ 0.89$ \\
 GC7 & 15 & 14 & $-0.54$ & $ 0.74$ \\
 GC8 & 12 &  9 & $-0.39$ & $ 0.60$ \\
 GC9 & 23 & 23 & $-0.99$ & $ 0.97$ \\
GC10 & 19 & 18 & $-0.67$ & $ 0.51$ \\
GC11 & 11 &  7 & $ 2.64$ & $ 0.83$ \\
GC12 & 22 & 15 & $-0.31$ & $ 0.07$ \\
GC13 & 20 & 18 & $-0.34$ & $ 0.71$ \\
GC14 & 18 & 16 & $-0.29$ & $ 0.93$ \\
GC15 &  8 &  5 & $ 0.13$ & $ 0.78$ \\
GC16 & 26 & 25 & $-0.52$ & $ 0.69$ \\
GC17 & 17 & 16 & $-0.16$ & $ 0.81$ \\
GC18 &  7 &  7 & $ 0.13$ & $ 0.79$ \\
GC19 & 21 & 20 & $-0.09$ & $ 0.93$ \\
GC20 & 19 & 11 & $-0.17$ & $ 0.70$ \\
GC21 & 10 &  4 & $-0.16$ & $ 0.88$ \\
GC22 & 23 & 20 & $-0.20$ & $ 0.92$ \\
GC23 & 21 & 16 & $-0.34$ & $ 0.57$ \\
GC24 &  5 &  4 & $ 0.13$ & $ 1.25$ \\
GC25 &  6 &  5 & $-0.26$ & $ 0.94$ \\
\hline
Total$^*$ & 405 &  348 \\
\hline
\end{tabular}
\end{center}
$^*$Duplicate detections in neighbouring fields are not counted.
\end{minipage}
\end{table}

\begin{table}
\begin{minipage}{70mm}
\begin{center}
\caption{
G01 variables whose counterparts were not found in our survey.
Superscripts ($a$--$d$) indicate our suggested explanations.
\label{tab:unmatchedGlass}}
\begin{tabular}{cl}
\hline
Field & Unmatched objects \\
\hline
 3 & 2389$^d$, 2832$^b$, 2834$^b$ \\
 5 & 59$^a$, 164$^a$, 2856$^a$ \\
 6 & 112$^d$ \\
 7 & 361$^d$ \\
 8 & 5$^a$, 31$^a$, 97$^b$ \\
 10 & 60$^b$ \\
 11 & 23$^a$, 27$^a$, 2449$^a$, 4503$^b$ \\
 12 & 2$^d$, 6$^a$, 11$^a$, 136$^a$, 228$^a$, 799$^d$, 1236$^d$ \\
 13 & 30$^a$, 73$^a$ \\
 14 & 150$^a$, 463$^a$ \\
 15 & 5$^a$, 10$^a$, 26$^a$ \\
 16 & 2993$^d$ \\
 17 & 118$^c$  \\
 19 & 685$^d$ \\
 20 & 11$^a$, 22$^a$, 46$^a$, 70$^a$, 99$^a$,
      116$^a$, 133$^b$, 522$^c$ \\
 21 & 6$^a$, 8$^a$, 17$^a$, 38$^a$, 185$^a$, 5719$^a$ \\
 22 & 31$^d$, 60$^a$, 100$^a$ \\
 23 & 5$^a$, 15$^c$, 42$^a$, 50$^a$, 114$^a$ \\
 24 & 29$^a$ \\
 25 & 7$^a$ \\
\hline
\end{tabular}
\end{center}
$^a$~The object is located out of our survey area.
$^b$~A probable counterpart exists in our image, but we did not detect
variability.
$^c$~Two or more stars are visible with approximately the expected brightness,
but we did not detect any corresponding variation.
$^d$~We found no clear counterpart at around the given coordinate.
\end{minipage}
\end{table}

\begin{figure}
\begin{center}
\begin{minipage}{75mm}
\begin{center}
\includegraphics[clip,width=0.85\hsize]{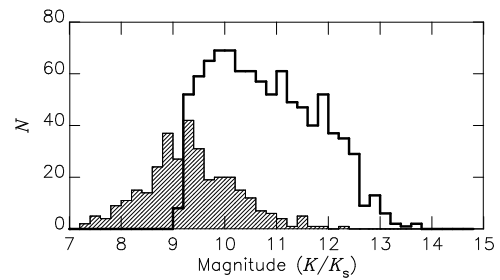}
\end{center}
\caption{
Histogram of $\Ks$ magnitudes which appear in our catalogue (the thick outline)
and $K$ magnitudes of the G01 variables (hatched region)
\label{fig:Mhist_GI}}
\end{minipage}
\end{center}
\end{figure}

We plotted a histogram of $K/\Ks$ mean magnitudes which appear
in both the G01 catalogue and ours in Fig.~\ref{fig:Mhist_GI}.
We used only good estimates of
$\Ks$ magnitudes for which $\Mf=0$ is given (see section~\ref{sec:mean}). 
Although we cannot directly compare the magnitudes obtained with different
filters, $K$ and $\Ks$, their difference is not expected to be large. It is
obvious that many bright sources are saturated in our data and that many
fainter ones were not detected in G01. With our better image quality, our
detection limit is deeper and we also discovered many new variables which
are brighter than the faintest G01 objects ($K\sim 12$),
as well as fainter variables.
Whilst many G01 variables were saturated in our $\Ks$-band images, a large
fraction of them were detected in the $J$- and/or $H$-bands. There are 348
matches between the two catalogues and none of these were saturated in our
$J$-band images. It should also be mentioned that saturation was not the
reason for the 56 unsuccessful matches (see Table~\ref{tab:unmatchedGlass}).

\subsection{Estimation of periods\label{sec:Pestimate}}

We used both our own data and the G01 catalogue
in order to estimate the periods of the detected variables.
First we carried out a period search of our monitoring data
using the least-squares method to fit sinusoidal curves.
We searched for periods between 100 and 800~d
fitting sinusoidal curves with minimum residuals.
Then we examined the light curve folded with the obtained period by eye.
We confirmed the periods for 483 variable stars.
The periods obtained by us are written as $\PI$.

\begin{figure}
\begin{center}
\begin{minipage}{75mm}
\begin{center}
\includegraphics[clip,width=0.85\hsize]{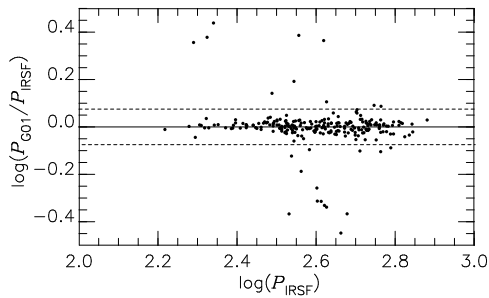}
\end{center}
\caption{
Comparison of $\PI$ and $\PG$. The $\log~(\PG/\PI)$
values are plotted against $\log~(\PI)$ whenever possible.
The dashed lines indicate the range of $\pm 0.075$~dex
within which we conclude $\PI$ and $\PG$ agree with each other.
\label{fig:Pcomp2}}
\end{minipage}
\end{center}
\end{figure}

There are 246 sources for which G01 obtained periods, $\PG$, among the
variables that have $\PI$ values. We have plotted $\log (\PG/\PI)$ values
against $\log (\PI)$ in Fig.~\ref{fig:Pcomp2}. The periods agree with each
other in most cases. The standard deviation of $\log (\PG/\PI)$ around zero
is 0.025, where we have included those within $\pm 0.1$~dex. This indicates
that the accuracies of the periods estimated by us and by G01 are within
0.025~dex in most cases. We conclude that $\PG$ and $\PI$ are consistent
with each other when their difference is less than 0.075~dex; this level is
indicated by the horizontal dashed lines in Fig.~\ref{fig:Pcomp2}. When the
values are consistent we take the period as $\PI$ in our catalogue. For the
25 objects where $\PG$ is not consistent with $\PI$ we also use $\PI$.
For the objects with counterparts in G01, we can combine their photometry
with ours to estimate the period. We fitted sinusoidal light curves
for both $K$-band (G01) and $\Ks$-band (ours) photometry;
means and amplitudes are independently obtained for two datasets
allowing differences in $K$- and $\Ks$-band,
but common periods and dates of maxima are used for the fitting.
These procedures allow us to carry out better period estimations
using the photometry with a long time coverage of 1994--2008.
We confirmed that thus obtained periods agree well with $\PI$
to within $\pm 0.01$~dex, and we replaced the $\PI$ whenever possible.

\begin{figure*}
\begin{minipage}{150mm}
\begin{center}
\includegraphics[clip,width=0.85\hsize]{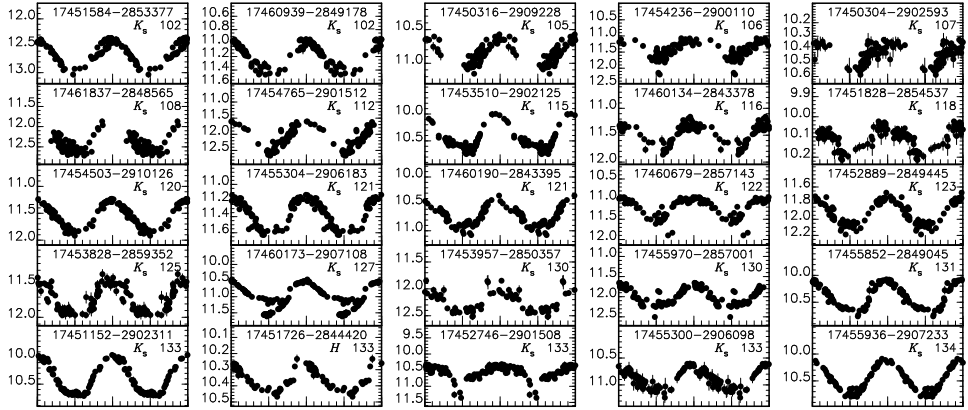}
\end{center}
\caption{
Examples of periodic light curves are plotted for
25 Miras with the shortest periods. In each panel we list
ID, period (d), and name of filter used for the plot.
Plots for all of those with given period are available as online material.
\label{fig:PLCs}}
\end{minipage}
\end{figure*}

There are 96 variables with $\PG$ among the variables whose periods
were not obtained by us. For 66 of them, we have confirmed that
$\PG$ can reasonably be used to explain the variations we detected. We have
adopted those $\PG$ in our catalogue. However,
$\PG$ does not fit the light curves for the remaining 30 variables. We use
$\Pf$ as defined in Table \ref{tab:Pflag} to indicate the status of
the periods. The numbers of variables with each $\Pf$ are also given in
Table \ref{tab:Pflag}. In summary,
we obtained periods for 549 variables ($\Pf=0$ to 3).
Light curves folded with the obtained periods are included in the online
material. Fig.~\ref{fig:PLCs} shows examples of the light curves for the
twenty-five cases having the shortest periods.

\begin{table}
\begin{minipage}{70mm}
\caption{
Definition of $\Pf$. Number of variables with each $\Pf$ is also given.
\label{tab:Pflag}}
\begin{center}
\begin{tabular}{ccp{55mm}}
\hline
$\Pf$ & $N_{\mathrm{var}}$ & Note \\
\hline
0 & 221 & $\PG$ and $\PI$ agree with each other to within 0.075~dex.
We adopt $\PI$. \\
1 & 25 & $\PG$ and $\PI$ disagree and we adopt $\PI$. \\
2 & 237 & $\PG$ is not given and we adopt $\PI$. \\
3 & 66 & $\PG$ fit our light curve but $\PI$ is not properly obtained. We adopt $\PG$. \\
4 & 30 & $\PG$ does not fit the IRSF light curve. No periodicity is found. \\
5 & 785 & $\PG$ is not given and no clear periodicity is found
with the IRSF data. \\
\hline
\end{tabular}
\end{center}
\end{minipage}
\end{table}

\subsection{Estimation of mean magnitudes\label{sec:mean}}

In this paper, for simplicity, we define our mean magnitude as the intensity
mean of the maximum and minimum of our measurements of a given star.
We can apply this definition, unlike Fourier means, regardless of whether or
not the light curve is periodic. Another possible definition is an intensity
mean of all the measurements, but this can be affected by uneven sampling.

In order to trust any photometric distance scale, it is important to obtain
magnitudes for both targets and calibrators in the same manner. In the case
of the PLR for pulsating stars, the definition of mean magnitude can have a
significant effect. For example, intensity means differ from magnitude
means. In our case, we will estimate mean magnitudes in the same manner as
%\citet{Ita-2004b}
\citeauthor{Ita-2004b}~(\citeyear{Ita-2004b}, hereafter I04)
because we will use their catalogue to calibrate our PLR (see
section~\ref{sec:LMCPLR}). I04 combined images at ten arbitrary epochs
before carrying out photometry, so that their magnitudes correspond to
intensity means of ten points. Because we have observations at more epochs,
approximately 90 in most cases, we should be able to get better estimates of
mean magnitudes. Nevertheless, it is necessary to be careful about any
systematic difference between our means and those in I04.

\begin{figure*}
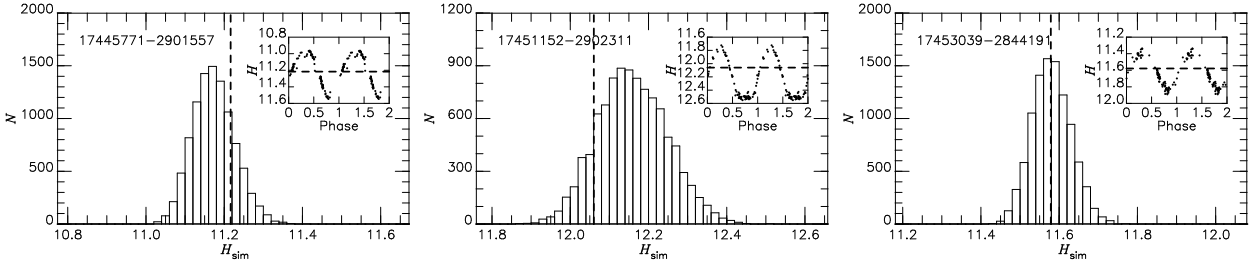

\begin{tabular}{ccc}
\begin{minipage}{55mm}
\begin{center}
\includegraphics[clip,width=0.98\hsize]{Fig8a.ps}
\end{center}
\end{minipage}
\begin{minipage}{55mm}
\begin{center}
\includegraphics[clip,width=0.98\hsize]{Fig8b.ps}
\end{center}
\end{minipage}
\begin{minipage}{55mm}
\begin{center}
\includegraphics[clip,width=0.98\hsize]{Fig8c.ps}
\end{center}
\end{minipage}
\end{tabular}
\caption{
Three examples of Monte-Carlo simulations of mean magnitudes from ten-epoch
data are shown. For each $H$-band light curve displayed in the small frame,
intensity means were repeatedly calculated for randomly defined groups of
ten epochs. The means obtained are plotted as a histogram, while the thick
dashed line indicates the intensity mean of maximum and minimum, as listed
in the catalogue.
\label{fig:simulate_Imean}}
\end{figure*}

We have performed Monte-Carlo simulations to assess the uncertainties of the
ten-epoch means and their systematic difference from ours, if any. First, we
randomly produced ten values between 0 and 1 as test phases, and for each of
them we obtained a magnitude from an interpolation between the closest
points on both sides of our folded light curve. Then, we took an intensity
mean of the magnitudes at the selected ten phases, which gives us a
simulated value for a ten-epoch mean. We repeated these procedures ten
thousand times for each light curve to get simulated magnitudes,
$m_\mathrm{sim}$.
Then we examined how the ten-epoch intensity mean varies, depending on the
randomly selected phases. We illustrate three examples of the simulations in
Fig.~\ref{fig:simulate_Imean}. The histograms indicate how the simulated
means, $H_\mathrm{sim}$, are distributed around the means of maximum and
minimum, which are indicated by the vertical dashed lines. The dispersions
$\sigma_\mathrm{sim}$ of the histograms suggest that a ten-epoch mean may
have a significant uncertainty as large as $\pm 0.1$~mag (e.g.\ the middle
panel of Fig.~\ref{fig:simulate_Imean}). Furthermore, the means of the
simulated magnitudes $\langle m_\mathrm{sim}\rangle$ do not always agree
with the vertical lines. Such  differences depend on the shapes of the
light curves.

\begin{figure}
\begin{center}
\begin{minipage}{75mm}
\begin{center}
\includegraphics[clip,width=0.85\hsize]{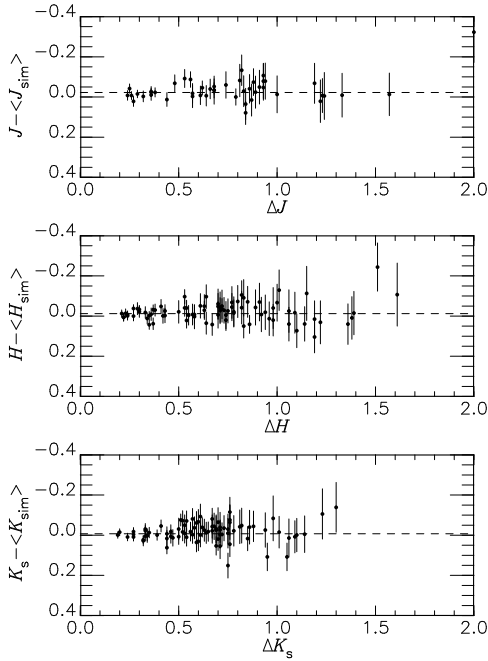}
\end{center}
\caption{
Expected differences between means obtained from observation, $m$,
and those from the Monte-Carlo simulations, $\langle m_\mathrm{sim}\rangle$.
The size of an error bar indicates the uncertainty in
$\langle m_\mathrm{sim}\rangle$.
\label{fig:comp_Imean}}
\end{minipage}
\end{center}
\end{figure}

We now examine whether the difference between I04 and our results introduces
a systematic offset or not. We consider only light curves which have good
coverage in phase for the Monte-Carlo simulations. We also exclude light
curves whose periods are not confirmed with our data (i.e.\ $\Pf$ in
section~\ref{sec:Pestimate} should be 0, 1, or 2). There are 48~($J$),
84~($H$), and 85~($\Ks$) such light curves without any gap larger than 0.2
in phase among those with $P=100$--$350$~d. We calculate differences between
the means from our observation and those from the simulations, i.e.\
$m-\langle m_\mathrm{sim}\rangle$, which are plotted against amplitudes in
Fig.~\ref{fig:comp_Imean}. The uncertainties $\sigma_\mathrm{sim}$ of a
ten-epoch mean inferred from the simulations are indicated by error bars.
One finds that both $|m-\langle m_\mathrm{sim}\rangle |$ and
$\sigma_\mathrm{sim}$ tend to be large for light curves with large
amplitude. The amplitudes of the examples presented in
Fig.~\ref{fig:simulate_Imean} are moderate among our sample. The weighted
mean of the differences is close to zero. This suggests that the
difference between our means and those in I04 is small, within $\pm
0.01$~mag or less.

We calculate the mean magnitudes, defined as intensity means of maximum and
minimum, in $\JHK$ whenever possible. However, we did not always detect
variables in all the three bands. Some variables are detected in $H$ and
$\Ks$ but are too faint in $J$, while some others are detected in $J$ and
$H$ but are too bright and saturated in $\Ks$-band. We define the
$\Mf$ as described in Table \ref{tab:Mflag} in order to give an idea of the
quality of the magnitude or a reason why there was no detection in a
particular band. In some cases, the number of detections is enough to
include the object in our list, i.e.\ $\Ndet \geq 0.5 \Nobs$, but a few
measurements are missing at around maximum or minimum for instrumental
reasons. We checked light curves for which we failed to get one or more
points. The phases of the failures were indicated in the plots and we
examined whether the unavailable points may have had a significant effect on
the mean magnitudes, based on the shapes of the light curves in a given
filter and in other filter(s) when available. Then we determined whether
the light curves are affected ($\Mf=4$--$6$) or not ($\Mf=0$). The
$\Mf$ for each variable has three digits between 0 and 7, each of which
denotes the state of (un)available magnitudes in the order of $J$, $H$,
and $\Ks$.

\begin{table*}
\begin{minipage}{160mm}
\caption{
Definition of $\Mf$. The number of variables with each $\Mf$ is also given
for $J$, $H$, and $\Ks$ magnitudes respectively.
\label{tab:Mflag}}
\begin{center}
\begin{tabular}{crrrp{110mm}}
\hline
$\Mf$ & $N_J$ & $N_H$ & $N_{\Ks}$ & Note \\
\hline
0 & 361 & 990 & 936 & The magnitude was obtained without any clear problems. \\
1 & 9 & 0 & 2 & We did not obtain the magnitude because the object often
falls out of the field of view. \\
2 & 0 & 18 & 222 & We did not obtain the magnitude because the object
is too bright. \\
3 & 833 & 178 & 4 & We did not obtain the magnitude because the object
is too faint. \\
4 & 15 & 35 & 21 & We obtained the mean magnitude, but it is affected by
 its location around the edge of the images. \\
5 & 2 & 19 & 116 & We obtained the mean magnitude, but it is affected by
 saturation or deviation from linear response.  \\
6 & 144 & 124 & 63 & We obtained the mean magnitude, but it is affected by
 the fact that the object was not detected in some cases
 because of its faintness. \\
\hline
\end{tabular}
\end{center}
\end{minipage}
\end{table*}

\subsection{Comparison with \citet{Ramirez-2008}}

\citet{Ramirez-2008} compiled a catalogue of 1,065,565 sources in the
$2.0^\circ \times 1.4^\circ$ region around the GC based on their {\it
Spitzer}/IRAC survey. About 60,000 objects lie within the field we observed.
They obtained magnitudes in four wave-passbands, i.e.\ $\IRACa, \IRACb,
\IRACc$, and $\IRACd$, so that we can investigate features of our objects at
longer wavelengths by making use of their results. We identified 1,233
counterparts in their catalogue with a tolerance of 0.5~arcsec. The
coordinates in the two catalogues agree very well, with standard deviations
of 0.13~arcsec in RA and Dec.

\subsection{The catalogue}

We present the first ten lines of our catalogue in Table~\ref{tab:CAT1}, as
an example. The entire list is available in machine-readable form in the
online version of this article. We include ID, mean magnitudes ($\JHK$),
peak-to-peak amplitudes ($\Delta J, \Delta H, \Delta \Ks$), and $\Mf$ (see
Table \ref{tab:Mflag}). The magnitudes and the amplitudes are listed as
99.99 and 0.00 respectively when we did not detect the object in a given
band. We also list the IDs of the sources in G01 and \citet{Ramirez-2008}
whenever counterparts were found. For those with period available, $P$
and $\Pf$ (see Table \ref{tab:Pflag}) are also indicated. We also present a
list of photometric measurements. The $\JHK$ magnitudes obtained for all the
catalogued variables are listed in one text file and each line includes
the ID number, the modified Julian date (MJD), and the $\JHK$ for a given star.
Table~\ref{tab:CAT2} shows the first ten lines as a sample.

\begin{table*}
\begin{minipage}{165mm}
\caption{Catalogue of variable stars.
After the numbering ID between 1 and 1364, the ID from RA and Dec (J2000.0) follows.
$\JHK$ mean magnitudes and peak-to-peak amplitudes are listed together with $\Mf$ (section \ref{sec:mean}). When the magnitudes and the amplitudes are unavailable, we put 99.99 and 0.00, respectively.
Counterparts of \citet{Ramirez-2008} and \citet{Glass-2001} are indicated whenever identified.
Also listed are periods (000 when not available) and \Pf (section \ref{sec:Pestimate}).
This is the first ten lines of the full catalogue which will be available in the online version.
\label{tab:CAT1}}
\begin{center}
\begin{tabular}{ccccccccccccc}
\hline
No. & ID & \multicolumn{3}{c}{Mean magnitude} & \multicolumn{3}{c}{Amplitude} & Mflag & \multicolumn{2}{c}{Counterpart} & Period & Pflag \\
  & & $J$ & $H$ & $\Ks$ & $\Delta J$ & $\Delta H$ & $\Delta \Ks$ & & R08 & G01 & & \\
\hline
0001 & 17445379-2857241 & 99.99 & 11.71 &  9.51 & 0.00 & 0.34 & 0.38 & 100 & 0401463 & -- & 000 & 5 \\
0002 & 17445418-2906303 & 99.99 & 12.50 &  9.71 & 0.00 & 0.94 & 0.71 & 300 & 0402528 & 22-166  & 339 & 0 \\
0003 & 17445435-2857415 & 99.99 & 11.39 &  9.15 & 0.00 & 0.88 & 0.33 & 105 & 0402897 & -- & 424 & 2 \\
0004 & 17445435-2904563 & 12.76 & 10.05 & 99.99 & 0.87 & 1.12 & 0.00 & 002 & 0402921 & 22-14   & 280 & 3 \\
0005 & 17445442-2856527 & 99.99 & 12.91 & 10.64 & 0.00 & 0.30 & 0.22 & 300 & 0403085 & -- & 000 & 5 \\
0006 & 17445452-2856168 & 99.99 & 13.35 & 11.09 & 0.00 & 0.62 & 0.44 & 300 & 0403313 & -- & 000 & 5 \\
0007 & 17445460-2852042 & 99.99 & 13.87 & 10.64 & 0.00 & 0.64 & 1.12 & 360 & 0403552 & -- & 000 & 5 \\
0008 & 17445463-2903587 & 99.99 & 13.50 & 10.13 & 0.00 & 1.68 & 1.48 & 360 & 0403617 & -- & 457 & 2 \\
0009 & 17445482-2903208 & 14.12 & 10.82 & 99.99 & 0.53 & 0.36 & 0.00 & 002 & 0404128 & -- & 000 & 5 \\
0010 & 17445482-2905486 & 16.13 & 11.87 &  9.46 & 0.58 & 0.54 & 0.60 & 000 & 0404117 & -- & 313 & 2 \\
\hline
\end{tabular}
\end{center}
\end{minipage}
\end{table*}

\begin{table}
\begin{minipage}{80mm}
\caption{The first 10 lines of the released table of light variation. This is a sample of the full version (116887 lines), which will be available in the online version of this journal.\label{tab:CAT2}}
\begin{center}
\begin{tabular}{ccccc}
\hline
No. & MJD & $J$ & $H$ & $\Ks$ \\
\hline
0001 & 52079.9757 & 99.99 & 11.62 &  9.52 \\
0001 & 52343.1326 & 99.99 & 11.62 &  9.68 \\
0001 & 53482.0590 & 99.99 & 11.76 &  9.54 \\
0001 & 53537.1116 & 99.99 & 11.79 &  9.55 \\
0001 & 53540.8221 & 99.99 & 11.83 &  9.58 \\
0001 & 53545.8849 & 99.99 & 11.84 &  9.64 \\
0001 & 53545.9694 & 99.99 & 11.85 &  9.57 \\
0001 & 53548.8259 & 99.99 & 11.81 &  9.54 \\
0001 & 53548.9591 & 99.99 & 11.84 &  9.56 \\
0001 & 53549.8893 & 99.99 & 11.84 &  9.62 \\
\hline
\end{tabular}
\end{center}
\end{minipage}
\end{table}

\section{Properties of catalogued variables\label{sec:Discuss1}}

\subsection{Magnitude and photometric flag}

Among the 1364 catalogued variables, we detected 522 variables in $J$, with
1168 in $H$ and 1136 in $\Ks$. If we restrict ourselves to those with $\Mf$
= 0, the numbers of variables are 361, 990, and 936 in $J$, $H$, and $\Ks$
respectively (Table \ref{tab:Mflag}). Table~\ref{tab:bands} indicates the
number and percentage for each combination of bands in which variables are
detected. For example, we detected 303 variables in all of the $\JHK$
filters while 178 were detected only in $\Ks$. If we consider only
magnitudes with $\Mf$ = 0, the above numbers reduce to 119 and 121,
respectively. Approximately 70 percent of the catalogued variables have both
$H$ and $\Ks$ magnitudes (22.2 percent with $\JHK$ and 48.0 percent with
$H\Ks$).

\begin{table}
\begin{minipage}{70mm}
\caption{Combinations of bands in which we detected variables.
The number and percentage are given for each combination.
The values in parentheses count only variables with
photometry flags of 0 for the given filters.
\label{tab:bands}}
\begin{center}
\begin{tabular}{crlrl}
\hline
Band combination & \multicolumn{2}{c}{Number} & \multicolumn{2}{c}{Percentage (\%)} \\
\hline
$J+H+\Ks$                       & 303 & (119) & 22.2 & (8.7)  \\
$\hspace{3.0mm}+H+\Ks$          & 655 & (485) & 48.0 & (35.6) \\
$J+\hspace{3.0mm}+\Ks$          &   0 & (0)   &  0.0 & (0.0)  \\
$J+H+\hspace{3.0mm}$            & 201 & (152) & 14.7 & (11.1) \\
$J\hspace{12.0mm}$              &  18 & (14)  &  1.3 & (1.0)  \\
$\hspace{6.0mm}H\hspace{6.0mm}$ &   9 & (7)   &  0.7 & (0.5)  \\
$\hspace{12.0mm}\Ks$            & 178 & (121) & 13.0 & (8.9)  \\
\hline
Total &  1364 & & 100.0 & \\
\hline
\end{tabular}
\end{center}
\end{minipage}
\end{table}

\begin{figure}
\begin{center}
\begin{minipage}{75mm}
\begin{center}
\includegraphics[clip,width=0.85\hsize]{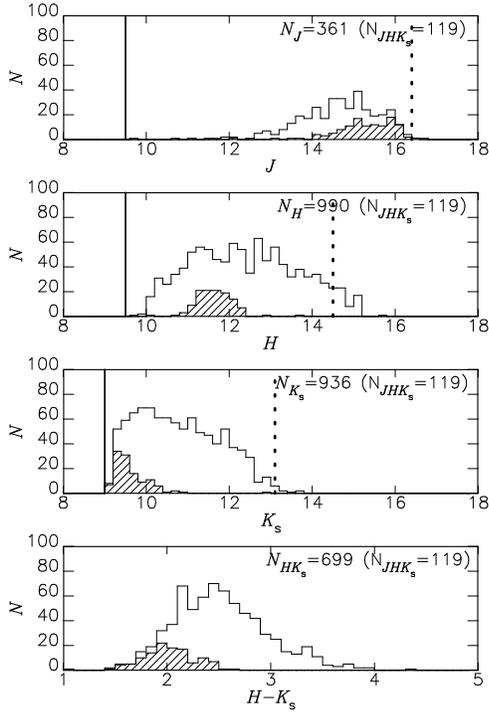}
\end{center}
\caption{
Histograms of obtained magnitudes, $J$, $H$ and $\Ks$ from top to bottom
and also that of the $H-\Ks$ colour.
Those for the objects with three-band magnitudes
are shaded.
\label{fig:MhistV}}
\end{minipage}
\end{center}
\end{figure}

We have plotted histograms of $J$, $H$, $\Ks$, and $H-\Ks$ in
Fig.~\ref{fig:MhistV}. Only magnitudes with $\Mf=0$ are considered. The
hatched regions correspond to the histograms for the variables whose $\JHK$
magnitudes are all available. Those with $\JHK$ magnitudes occupy
different
ranges in the three panels. For $J$- and
$H$-band magnitudes, the brightest objects tend to lack $\Ks$-band
magnitudes (see the regions of $J<14$ and $H<11$). This is because the
corresponding objects are too bright in $\Ks$. In contrast, $J$-band
magnitudes are not available for a large number of objects which are faint
in $H$ and $\Ks$ ($H>12.5$ and $\Ks>10.5$). This is because these objects
are too faint in $J$, predominantly because they suffer from strong
interstellar extinction. The bottom panel shows that only relatively blue
variables have all the $\JHK$ magnitudes, supporting the conclusions above.

\subsection{Colour-magnitude and colour-colour diagram\label{sec:CMDCCD}}

\begin{figure}
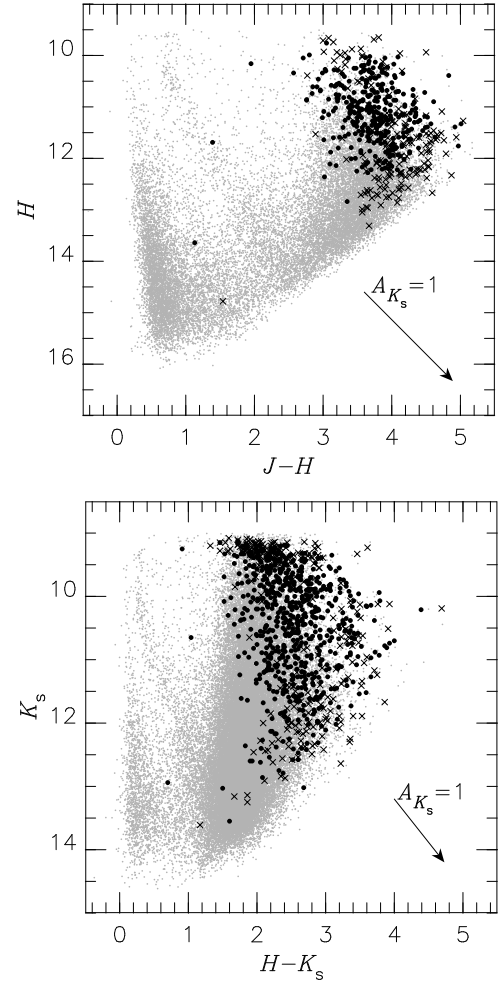

\begin{center}
\begin{minipage}{75mm}
\begin{center}
\includegraphics[clip,width=0.85\hsize]{Fig11a.ps}
\vspace{3mm}

\includegraphics[clip,width=0.85\hsize]{Fig11b.ps}
\end{center}
\caption{
Colour-magnitude diagrams, $(J-H)$-$H$ in the top panel
and $(H-\Ks)$-$\Ks$ in the bottom.
The black points show the distribution of the catalogued variables
with $\Mf=0$ while the black crosses include those with $\Mf=4$--$6$.
Grey dots show all the detected non-variable sources.
The arrow indicates the reddening vector corresponding to $\AK=1$~mag.
\label{fig:CMD}}
\end{minipage}
\end{center}
\end{figure}

In Fig.~\ref{fig:CMD}, we present colour-magnitude diagrams for our
catalogued variables (black points for those with $\Mf=0$ and black crosses
for those with $\Mf=4$--$6$) and non-variables (grey dots). The arrows
indicate the reddening vector directions, with lengths corresponding to
$\AK=1$~mag. It is apparent that the objects with strong extinction are
often fainter than the detection limit, especially in the $J$-band.

\begin{figure}
\begin{center}
\begin{minipage}{75mm}
\begin{center}
\includegraphics[clip,width=0.85\hsize]{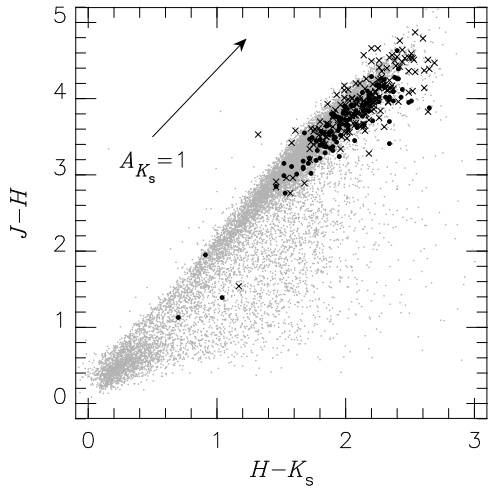}
\end{center}
\caption{
Colour-colour diagram, $(H-\Ks)$ -- $(J-H)$.
Symbols are the same as in Fig.~\ref{fig:CMD}.
The arrow indicates the reddening vector corresponding to $\AK=1$~mag.
\label{fig:CCD}}
\end{minipage}
\end{center}
\end{figure}

\begin{figure}
\begin{center}
\begin{minipage}{75mm}
\begin{center}
\includegraphics[clip,width=0.95\hsize]{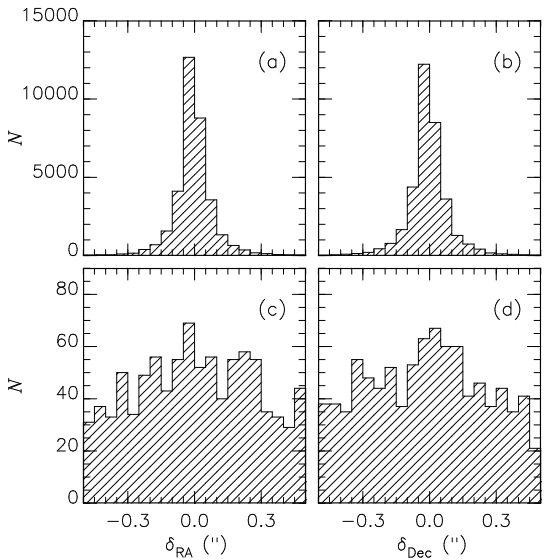}
\end{center}
\caption{
Histograms of the difference between positions in $J$ and $\Ks$ images.
The upper panels (a) and (b) are for objects with normal locations
in the colour-colour diagram, 
while the lower panels (c) and (d) are for objects with peculiar locations,
i.e.\  $J-H<1.767(H-\Ks)-1$.
\label{fig:dxdy}}
\end{minipage}
\end{center}
\end{figure}

We also plot the colour-colour diagram in Fig.~\ref{fig:CCD}. Due to large
and variable extinction, the distribution of sources is elongated along the
direction of the reddening vector. There is a group of stars which scatter
towards the right side of the main diagonal sequence. We inspected these
stars on the images and found that tiny displacements of centroid are
recognizable among the $\JHK$ images for at least some of them. In
Fig.~\ref{fig:dxdy}, we plot histograms of the differences between
positions, both RA and Dec, obtained in $J$ and $\Ks$ images for individual
sources. The panels (a) and (b) include the objects belonging to the main
group and the distributions resemble Gaussian functions centred on zero as
expected. On the other hand, the panels (c) and (d) include objects with
peculiar colours, i.e.\ $J-H<1.767(H-\Ks)-1.0$, and show totally different
distributions from those in (a) and (b). There is no sharp peak at zero and
the differences distribute rather uniformly. This clearly suggests that
different stars having almost the same positions were incorrectly combined
to produce sources with peculiar colours. In other words, different stars
may contribute to the fluxes we measured in the images in different filters:
a bluer star to the $J$-band flux and a redder star to $\Ks$-band. If we
combine the flux of a foreground star on the main sequence and that of a
reddened giant, for example, the combined source will be located on the
right side of the main distribution in Fig.~\ref{fig:CCD}. The objects with
$J-H<1.767 (H-\Ks) -1.0$ constitute not more than 3 percent of the number of
sources with all three $\JHK$ magnitudes, so that they hardly affect any of
our results.

\subsection{Period distribution\label{sec:period}}

\begin{figure}
\begin{center}
\begin{minipage}{75mm}
\begin{center}
\includegraphics[clip,width=0.85\hsize]{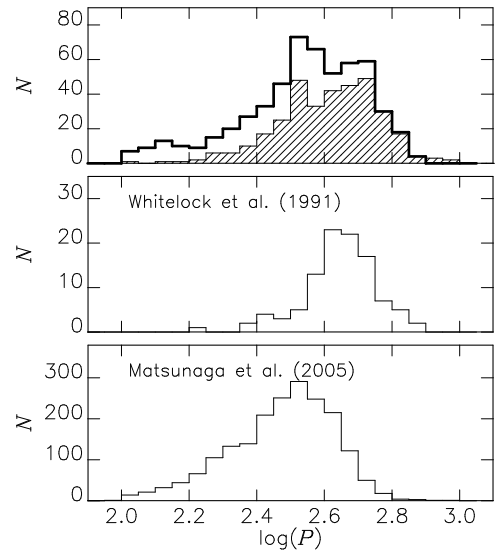}
\end{center}
\caption{
Histograms of periods. In the top panel, the thick line shows the histogram
of the periods adopted in our catalogue, 
while the shaded area indicates that of $\PG$ for the G01 variables
with IRSF counterpart. In the middle and bottom panels shown are
the period distributions reported in \citet{Whitelock-1991} and
M05 respectively.
\label{fig:Phist}}
\end{minipage}
\end{center}
\end{figure}

We have plotted period histograms for the variables in our catalogue and
also in a few others (Fig.~\ref{fig:Phist}). Compared with the sample in G01
(the shaded area) in the top panel, we have added significantly more
variables with periods (see the thick line). The other panels show the
histograms for the variables in
%% \citeauthor{Whitelock-1991}~(\citeyear{Whitelock-1991},
Whitelock, Feast \& Catchpole (1991, middle) and in
M05 (bottom). Their
objects are located away from the GC. The sample in
\citet{Whitelock-1991} was selected from the {\it IRAS} point-source
catalogue, while M05 investigated variables found in the
OGLE variability catalogue \citep{Wozniak-2002}.
Because of the selection bias caused by the different wavelengths
used in the surveys,
the period distributions in these two early studies are different
from each other as discussed in M05.
The former has variables with $P>600$~d, while the latter has a
more significant tail towards the short period end.
In contrast, the histogram for our catalogue includes
both the long-period variables and the short-period tail.
It is suggested that our catalogue contains a more comprehensive
sample of Miras than previous ones.

\subsection{Amplitude distribution\label{sec:amplitude}}

\begin{figure}
\begin{center}
\begin{minipage}{75mm}
\begin{center}
\includegraphics[clip,width=0.85\hsize]{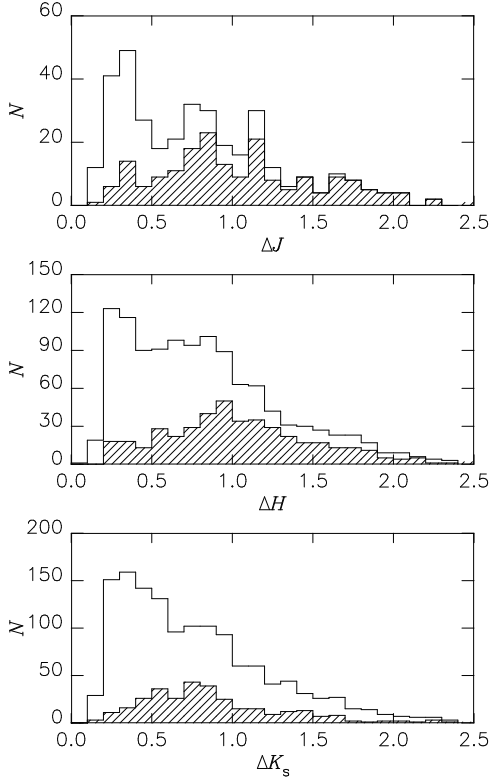}
\end{center}
\caption{
Histogram of amplitudes: $\Delta J$, $\Delta H$ and $\Delta \Ks$
from the top panel to the bottom.
The outline includes
all the variables, while the shaded area indicates
those with period.
\label{fig:Amphist}}
\end{minipage}
\end{center}
\end{figure}

Fig.~\ref{fig:Amphist} shows histograms of peak-to-peak amplitudes in $\JHK$.
The hatched region corresponds to the variables with periods (191 in
$J$, 432 in $H$ and 319 in $\Ks$), while the outline includes all the
variables with $\Mf=0$ detected in each filter. The histogram of $\Delta J$
shows separated peaks: one for small-amplitude variables ($\Delta J < 0.5$)
and the other(s) for large-amplitude variables. Such a distribution is also
seen in the histogram of optical amplitudes (e.g.\ see fig.\ 4 and 6 in
\citealt{Glass-2003}). Many of the large-amplitude variables have been found
to be periodic as expected for Miras, whilst we failed to obtain periods for
many of the small-amplitude ones. The separation gets less conspicuous as
the wavelength increases. The hatched region in the $\Delta\Ks$ panel
resembles the distribution for Miras reported by
\citeauthor{Glass-1995}~(\citeyear{Glass-1995}; see their fig.\ 3).

It is traditional to separate the Miras from the other long-period
variables such as semi-regulars by their large visible amplitudes,
i.e.\  $\Delta V \geq 2.5$~mag. As we will see in section~\ref{sec:LMCPLR}, we
select periodic variables with $\Delta I > 1$~mag among the LMC variables
reported in I04 to calibrate the PLR. However, we do not have $V$- or
$I$-band amplitudes for our objects.
Most Miras have amplitudes larger than
0.4~mag in $K$ (e.g.\ see %% \citet{Whitelock-2000}),
Whitelock, Marang \& Feast, 2000), but the threshold in the
near-IR amplitudes is not well established.  It is even unclear if one can
pick up the same group of Miras by using observables at different
wavelengths. There are some periodic variables with relatively small
amplitude. For example, 17451726-2844420 ($P=133$) and 17462476-2902241
($P=177$) have rather small amplitudes ($\Delta J < 0.3$~mag), but their
light curves show reasonably periodic variations
(Fig.~\ref{fig:PLCs}). These objects may be SRa type, which can be as
periodic as Miras but are distinguished by having relatively small
amplitudes. However, some recent studies have raised questions on the
distinction between the Miras and at least some of the semi-regulars.
%% \citet{Lebzelter-2002},
Lebzelter, Schultheis \& Melchior (2002),
for example, showed that a sizeable number of
small-amplitude variables are as regular as Miras.

The most important requirement in our selection is that our distance
indicators follow the PLR of Miras. There are several sequences in the
well-known period-luminosity diagram of pulsating red giants
(\citealt{Wood-1999}; \citealt{Ita-2004a}).
Nevertheless, it is known that Miras if
properly chosen are exclusively found on the sequence C, which makes them
valuable distance indicators. Unfortunately, it is currently impossible to
tell which sequence a given small-amplitude variable is located on
as \citet{Feast-2004} pointed out.
Some small-amplitude variables coexist with Miras on the sequence C.
We here consider a periodic variable to be a Mira if its 
$\Delta J$, $\Delta H$, or $\Delta \Ks$, whichever may be available,
is not less than 0.4~mag. Among our objects with periods, the number of
such small-amplitude variables excluded is not large, about 10 percent.
As will be
discussed in section \ref{sec:SmallAmp}, at least some of the
small-amplitude variables seem to be on a different sequence from that of
Miras in the P-L diagram.

\begin{figure}
\begin{center}
\begin{minipage}{75mm}
\begin{center}
\includegraphics[clip,width=0.85\hsize]{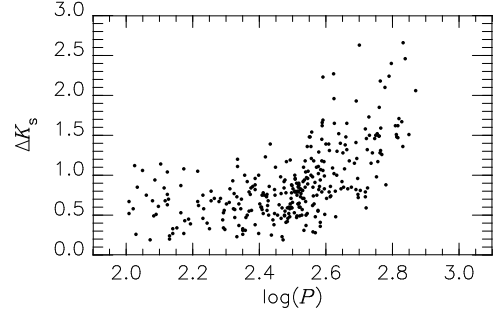}
\end{center}
\caption{
Amplitudes in the $\Ks$-band plotted against periods.
\label{fig:PAmp}}
\end{minipage}
\end{center}
\end{figure}

Fig.~\ref{fig:PAmp} shows $\Ks$-band amplitude plotted against period.
One finds that the amplitudes do not exceed 1.5~mag for
those with $P=100$--$350$~d, which we will use as distance indicators.
This property does not change in the cases of $\Delta J$ and $\Delta H$.

\subsection{Mid-IR properties}

\begin{figure}
\begin{center}
\begin{minipage}{75mm}
\begin{center}
\includegraphics[clip,width=0.85\hsize]{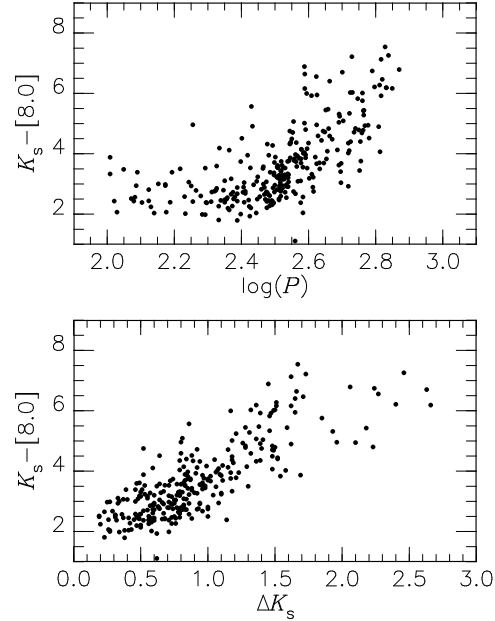}
\caption{
The $\Ks-\IRACd$ colours are plotted against
the periods (top) and the amplitudes (bottom) for periodic variables.
\label{fig:CvsPA}}
\end{center}
\end{minipage}
\end{center}
\end{figure}

Using the counterparts identified in the catalogue by \citet{Ramirez-2008},
we can study the mid-IR properties of our variables. We have plotted
$\Ks-\IRACd$ colour against period and amplitude for 282 Miras with $\Mf=0$
in $\Ks$. In the top panel of Fig.~\ref{fig:CvsPA}, the mid-IR colour
becomes large at the long-period end, $P > 350$~d, similar to the amplitude
trend illustrated in Fig.~\ref{fig:PAmp}. The lower panel of
Fig.~\ref{fig:CvsPA} shows that the variables with large enhancement in
$\Ks-\IRACd$ are, in fact, associated with large amplitudes. We also
examined similar plots using $\IRACa, \IRACb$ and $\IRACc$ magnitudes
instead of $\IRACd$, and found similar trends. A large excess in the mid IR
is expected for a thick circumstellar dust shell produced through mass-loss. 
This behaviour is well known from early observations (e.g.\
\citealt{Whitelock-1994}). Detailed study of the mid-IR properties and the
mass-loss process is outside the scope of this paper. Please note that no
correction has been made for interstellar extinction and also that the
mid-IR magnitudes from \citet{Ramirez-2008} were obtained from a
single-epoch survey made with {\it Spitzer Space telescope}.

\section{Distance and Reddening\label{sec:Discuss2}}

\subsection{Relation from the LMC\label{sec:LMCPLR}}

\begin{figure}
\begin{center}
\begin{minipage}{75mm}
\begin{center}
\includegraphics[clip,width=0.85\hsize]{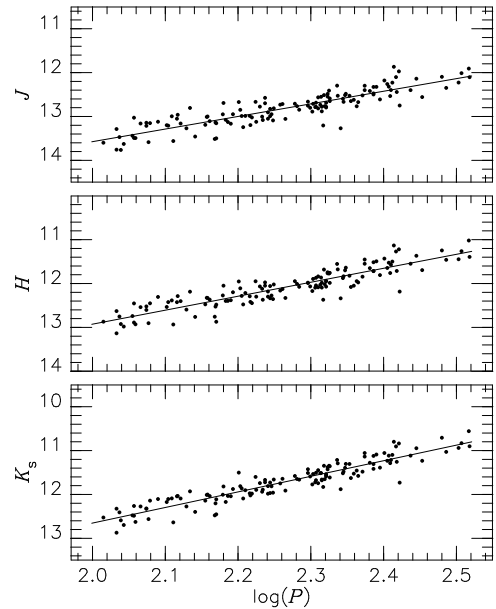}
\caption{
Period-luminosity relation for Miras in the LMC,
plotted for 134 O-rich Miras selected from the catalogue of I04
(see text for the selection criteria).
\label{fig:LMCPLR}}
\end{center}
\end{minipage}
\end{center}
\end{figure}

In this paper we make use of the PLR for Miras in the LMC in order to make
distance estimations. I04 published a near-IR catalogue of LMC variables
which were detected in the OGLE-II survey by %% \citet{Udalski-1997}.
Udalski, Kubiak \& Szyma\'{n}ski (1997). We select
134 Miras under the following criteria,
\begin{itemize}
\item $\Delta I > 1$,
\item $J-\Ks < 1.4$,
\item $\Ks < 13$,
\item $P \leq 350$,
\end{itemize}
as calibrators. The colour criterion is used to exclude carbon-rich stars.
Miras in the Galactic bulge are predominantly oxygen-rich (see
%% \citealt{Schultheis-2004},
Schultheis, Glass \& Cioni, 2004, for example). The last criterion is used to
exclude objects with thick circumstellar dust and hot-bottom-burning stars
(\citealt{Whitelock-2003}; \citealt{Glass-2003}). The near-IR observations
for the I04 catalogue also made use of IRSF/SIRIUS. Although I04
converted their magnitudes into the system of the Las Campanas Observatory
(LCO), we have converted them back into the natural system of the
IRSF/SIRIUS based on the transformation they used.
Fig.~\ref{fig:LMCPLR} plots the $\JHK$ magnitudes against periods.
The period distribution of the LMC calibrators is rather uniform
except the longest period range ($\log P >2.44$). In this range
more carbon-rich variables are found in the LMC than oxygen-rich ones
\citep{Groenewegen-2004}. This situation is different from that of the
Galactic bulge. Nevertheless, the effect of the period distribution
in determining the PLR
is small if any. There has been no report to offer evidence of
a change in the slope of the PLR.

The PLR can be determined to be:
\begin{eqnarray}
M_J = -2.875 (\log P -2.3) - 5.798, \label{eq:PLRJ} \\
M_H = -3.186 (\log P -2.3) - 6.508, \label{eq:PLRH} \\
\MK = -3.555 (\log P -2.3) - 6.883, \label{eq:PLRK}
\end{eqnarray}
with residual standard deviations of 0.19, 0.19 and 0.17~mag, respectively.
The distance of the LMC is assumed to be 18.45~mag (see section
\ref{sec:ErrorSize} for its uncertainty) and extinction corrections of
$A_J=0.06$, $A_H=0.03$ and $\AK=0.02$ have been applied. The residual
scatter is least in $\Ks$, but the differences are small. This indicates
that the PLR in $J$ and $H$ can also be useful for distance estimation.
More importantly, one can estimate the amount of interstellar extinction
together with the distance by using the PLR in two or more wavebands.
The residual scatter is slightly larger than that in \citet{Feast-1989} for
the $\log P$-$\Ks$ relation while those around the PLRs in $J$ and $H$
are similar. The relatively large scatter is presumably because the magnitudes
in I04 were obtained as intensity means based on ten-epoch monitorings. The
simulation presented in section~\ref{sec:mean} suggests that the ten-epoch
mean can have an uncertainty of $\pm 0.1$~mag compared to a well determined
value. One can expect that the scatter reduces to that reported in
\citet{Feast-1989} after mean magnitudes are properly obtained.

\begin{figure}
\begin{center}
\begin{minipage}{75mm}
\begin{center}
\includegraphics[clip,width=0.85\hsize]{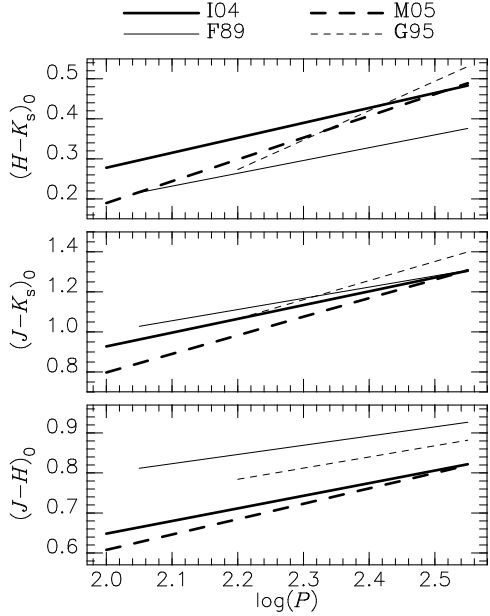}
\caption{
Period-colour relations for Miras obtained in earlier papers. Line styles
and widths for different samples are indicated at the top:
F89--\citet{Feast-1989}, G95--\citet{Glass-1995}, I04--\citet{Ita-2004b},
and M05--\citet{Matsunaga-2005}. The projection of each line on the
horizontal axis indicates the period range of the samples used to derive the
relation.
\label{fig:compPCR}}
\end{center}
\end{minipage}
\end{center}
\end{figure}

It is necessary to call attention to possible differences between Miras in
the LMC and those around the GC. It is suspected that the PLR of Miras may
be dependent on stellar parameters such as metallicity. Readers are referred
to \citet{Whitelock-2008} for a review of the PLRs in different
environments. On the other hand, a difference in photometric systems can
also have a large effect. Fig.~\ref{fig:compPCR} plots period-colour
relation (PCR) for the Miras in the LMC (\citealt{Feast-1989}; I04) and
those in the Galactic bulge (\citealt{Glass-1995}; M05).
For the relation of
\citet{Glass-1995}, we calculated the PCR by using the samples with period
shorter than 350~d after two objects with extreme colours were
omitted. Lines are drawn for a period range of samples in each paper; the
shortest period in \citet{Glass-1995} is around
$\log P=2.2$, while there are several objects with period close to
100~d in I04 and M05.
\citet{Feast-1989} and \citet{Glass-1995} used the Mk~{\small III} photometer
attached to the 1.9-m telescope at SAAO Sutherland to observe the Miras in
the LMC and those in the Galactic bulge towards the Sgr~I window
respectively. Observations by I04 were conducted by using the IRSF/SIRIUS
also at Sutherland, while M05 used
$\JHK$ magnitudes taken from the 2MASS point-source catalogue
\citep{Skrutskie-2006}. Although transformations between these systems are
available for normal types of stars, it is highly uncertain whether they are
applicable to Miras which have strong absorption due to H$_2$O and other
molecules. Based on Fig.~\ref{fig:compPCR}, it is difficult to discuss any
population effect that may exist between the Miras in the two populations.
Especially in the $\log P$-$(J-H)_0$
diagram, the difference between the instruments seems to have a larger
effect than any possible population differences. We will therefore use the
PLR, eqs.\ \ref{eq:PLRJ}--\ref{eq:PLRK}, obtained from the observations by
I04. The possible population effect will be discussed again in
section~\ref{sec:ErrorSize}.

\subsection{Estimation of $\mu _0$ and $\AK$\label{sec:Estimation}}

We can get the apparent distance modulus for an individual Mira of known
period using the PLR (eqs.~\ref{eq:PLRJ}--\ref{eq:PLRK}) if a mean magnitude
is available in a given filter. The apparent distance moduli in $\JHK$ are
related to the true distance modulus $\mu_0$ and the extinction $\AK$ as
follows:
\begin{eqnarray}
\mu_0 + r_J \AK = J - M_J, \label{eq:PLReqJ} \\ 
\mu_0 + r_H \AK = H - M_H, \label{eq:PLReqH} \\
\mu_0 + \AK = \Ks - \MK, \label{eq:PLReqK} 
\end{eqnarray}
where $r_J$ and $r_H$ are defined as $A_J/\AK$ and $A_H/\AK$ respectively.
The extinction coefficients have been obtained
by \cite{Nishiyama-2006a},
\begin{equation}
r_J = 3.021, ~r_H = 1.731, \label{eq:Reddening}
\end{equation}
using the same instrument as ours. When a period and mean magnitudes in two
or more filters are available, we can solve for the two unknown quantities
$\DMAK$. For a Mira with all three $\JHK$ magnitudes available, we can make
three estimates of $\DMAK$ by using the three different pairs of filters. In
order to specify the band pair used to derive a particular $\DMAK$, we use
the band names as superscripts, e.g. $\mu_0^{~HK}$ and $\AK^{~HK}$. For
example, the solution of eqs.\ (\ref{eq:PLReqH}) and (\ref{eq:PLReqK}) can
be written as follows:
\begin{eqnarray}
\AK^{~HK} = \frac{H-\Ks}{r_H-1} - \frac{M_H-\MK}{r_H-1},
  \label{eq:solutionAHK}\\
\mu_0^{~HK} = \frac{r_H\Ks-H}{r_H-1} - \frac{r_H\MK-M_H}{r_H-1}.
  \label{eq:solutionMHK}
\end{eqnarray}

\begin{table}
\begin{minipage}{80mm}
\caption{Extinctions $\AK$ and distance moduli $\mu_0$ estimated for Miras.
Three estimates from different pairs of filter bands are listed whenever available.
We put 99.99 if they are not available.
The full version will be available in the online journal.
\label{tab:CAT3}}
\begin{center}
\begin{tabular}{cccccccccccc}
\hline
No. & $\AK^{~HK}$ & $\mu_0^{~JK}$ & $\AK^{~JK}$ & $\mu_0^{~HK}$ & $\AK^{~JH}$ & $\mu_0^{~JH}$ \\
\hline
0002 &  3.19 & 14.22 & 99.99 & 99.99 & 99.99 & 99.99\\
0004 & 99.99 & 99.99 & 99.99 & 99.99 &  1.52 & 14.40\\
0010 &  2.69 & 14.35 &  2.70 & 14.34 &  2.71 & 14.32\\
0016 &  4.47 & 14.07 & 99.99 & 99.99 & 99.99 & 99.99\\
0023 &  1.71 & 14.96 &  1.82 & 14.84 &  1.89 & 14.65\\
0034 &  4.05 & 14.30 & 99.99 & 99.99 & 99.99 & 99.99\\
0035 &  2.69 & 14.87 & 99.99 & 99.99 & 99.99 & 99.99\\
0039 & 99.99 & 99.99 & 99.99 & 99.99 &  1.59 & 14.68\\
0040 &  2.02 & 14.61 &  2.03 & 14.61 &  2.03 & 14.60\\
0074 & 99.99 & 99.99 & 99.99 & 99.99 &  2.00 & 14.46\\
\hline
\end{tabular}
\end{center}
\end{minipage}
\end{table}

\begin{figure*}
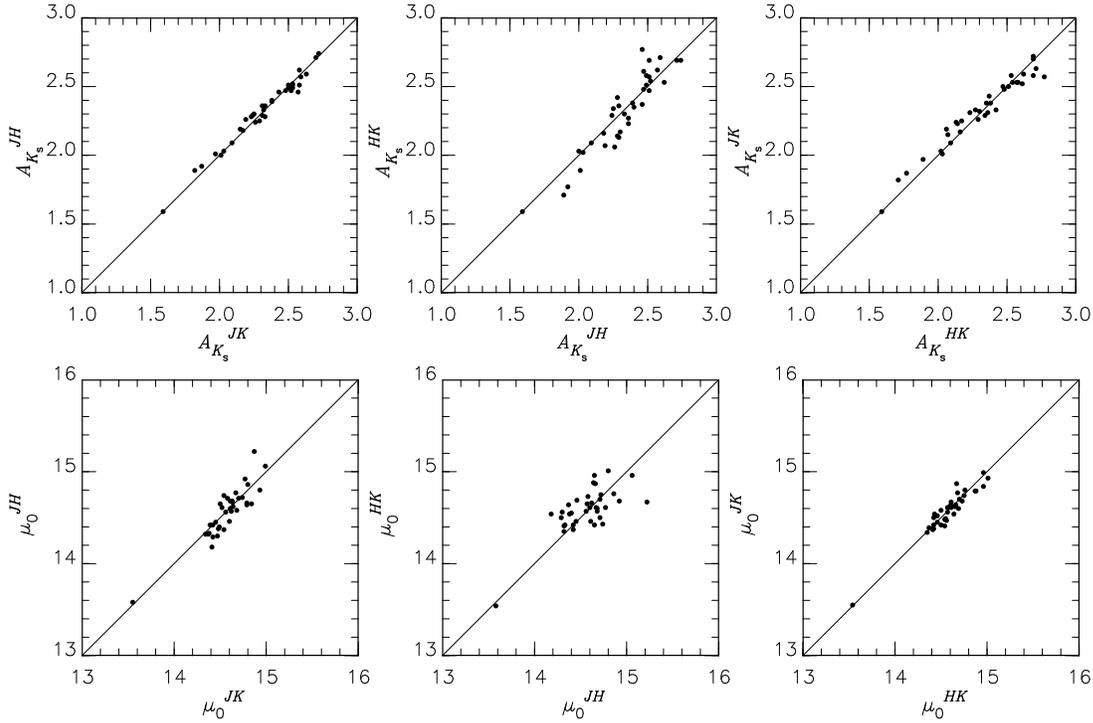

\begin{minipage}{160mm}
\begin{center}
\includegraphics[clip,width=0.90\hsize]{Fig20a.ps}
\includegraphics[clip,width=0.90\hsize]{Fig20b.ps}
\caption{
In the above three panels, we compare $\AK$ values obtained with
magnitudes in different pairs of filters.
Comparisons for $\mu_0$ values are also plotted in the lower panels. 
\label{fig:DMAKcomp}}
\end{center}
\end{minipage}
\end{figure*}

We have made distance estimations using only Miras with $P \leq 350$~d,
amplitudes not smaller than 0.4~mag and at least two mean
magnitudes having $\Mf=0$. There are 175 Miras fulfilling these
requirements; the numbers for the three pairs of filters, being 143
for $(H, \Ks)$, 37 for $(J, \Ks)$ and 69 for $(J, H)$.
Table~\ref{tab:CAT3} lists
the $\DMAK$ values found. For the 37 Miras which have all three $\JHK$ means,
we were able to obtain three estimates of $\DMAK$.
Ideally, the values found from the different pairs should agree with each
other. The $\DMAK$ values are mutually compared in Fig.~\ref{fig:DMAKcomp}.
The points are distributed around the diagonals but show some scatter. 

In the following, we estimate the uncertainties in solving
$\DMAK$ and show that the scattering in Fig.~\ref{fig:DMAKcomp}
can be explained quantitatively.
The solutions (\ref{eq:solutionAHK}), (\ref{eq:solutionMHK}), and similar
ones for other pairs of bands are affected by the uncertainties in our
photometry ($\JHK$) and in the predictions from the PLR ($M_J, M_H,
\MK$). The prediction of absolute magnitudes based on the PLR cannot be
totally accurate because there is scatter around each relation as seen in
Fig.~\ref{fig:LMCPLR}. The scatter can be as large as 0.19~mag, but the
deviations are not mutually independent. For the 134 LMC Miras in
Fig.~\ref{fig:LMCPLR}, the deviations from the PLR in $J$, $H$ and $\Ks$
are strongly correlated, which should be borne in mind when estimating
the uncertainty caused by the width of the PLR. The uncertainties in the
$\JHK$ magnitudes are considered to be mutually independent, and we estimate
that the errors in the mean magnitudes of $\JHK$ are 0.02~mag. We calculated
expected uncertainties in the differences between estimates of $\AK$ or
$\mu_0$, such as $\AK^{~JH}-\AK^{~JK}$ and $\mu_0^{~JH}-\mu_0^{~JK}$. The
results are listed in Table~\ref{tab:DMAKerror} together with the scatters
actually obtained in our results. The predicted uncertainties agree with the
actual values well. The uncertainties in the estimated periods can also
affect the estimation of $\DMAK$. However, they produce errors of
almost the same size for different band pairs,
and they compensate each other when we take the differences
listed in Table~\ref{tab:DMAKerror}. The
extinction coefficients $r_J$ and $r_H$
averaged over the GC region
are reported to have uncertainties
of about 1 percent \citep{Nishiyama-2006a}. From eqs.~(\ref{eq:solutionAHK})
and (\ref{eq:solutionMHK}), the resultant errors in $\AK^{~HK}$ and
$\mu_0^{~HK}$ are 0.06 and 0.07~mag, respectively. However,
%% the effects due to period uncertainty
they 
do not affect the differences between the results from
each colour combination. We therefore conclude that the scatters in
Fig.~\ref{fig:DMAKcomp} are adequately explained by the expected errors in
our method.

\begin{table}
\begin{minipage}{75mm}
\caption{
Errors in some values related to the estimation of $\DMAK$. The predicted
uncertainties are based on
eqs.~(\ref{eq:solutionAHK})--(\ref{eq:solutionMHK}) and similar solutions.
The parts related to observational magnitudes ($\JHK$) and those related to
the PLR predictions are individually listed together with the totals. The
scatters found in Fig.~\ref{fig:DMAKcomp} are listed in the last column. The
uncertainties in the periods affect the values given in the last two lines,
but they are not included here. They have almost no effect on the other
six lines (see text).
\label{tab:DMAKerror}}
\begin{center}
\begin{tabular}{ccccc}
\hline
Value & \multicolumn{3}{c}{Predicted uncertainty} & Observed \\
 & {\it Obs.} & {\it PLR} & {\it Total} & scatter \\
\hline
$\AK^{~JH}-\AK^{~JK}$     & 0.019 & 0.038 & 0.042 & 0.040 \\
$\AK^{~HK}-\AK^{~JH}$     & 0.053 & 0.105 & 0.118 & 0.110 \\
$\AK^{~JK}-\AK^{~HK}$     & 0.034 & 0.067 & 0.075 & 0.070 \\
$\mu_0^{~JH}-\mu_0^{~JK}$ & 0.038 & 0.111 & 0.117 & 0.123 \\
$\mu_0^{~HK}-\mu_0^{~JH}$ & 0.092 & 0.182 & 0.203 & 0.192 \\
$\mu_0^{~JK}-\mu_0^{~HK}$ & 0.034 & 0.067 & 0.075 & 0.070 \\
\hline
$\AK^{~HK}$               & 0.039 & 0.082 & 0.091 &       \\
$\mu_0^{~HK}$             & 0.039 & 0.179 & 0.183 &       \\
\hline
\end{tabular}
\end{center}
\end{minipage}
\end{table}

\citet{Glass-1995} suggested that the Miras in the Sgr~I field
have larger $(H-K)_0$ and smaller $(J-H)_0$ than those in the LMC,
at a given period, as can be seen in Fig.~\ref{fig:compPCR}.
If such a difference exists, our estimates would give
larger $\AK^{HK}$ and smaller $\AK^{JH}$ values
for the bulge Miras. In effect, in \citet{Glass-1995}
the estimated $A_{K}^{HK}$ values are systematically larger than
the $A_{K}^{JH}$
by 0.1--0.2~mag, depending on the period range considered.
We did not, however, find such a shift in Fig.~\ref{fig:DMAKcomp}.
It is therefore
suggested that the PCRs of the Miras in the LMC and the bulge
are similar to each other in the IRSF/SIRIUS system.
This may be confirmed in the future from IRSF photometry of the Miras
in the Sgr~I field and NGC~6522.

\begin{figure}
\begin{center}
\begin{minipage}{75mm}
\begin{center}
\includegraphics[clip,width=0.85\hsize]{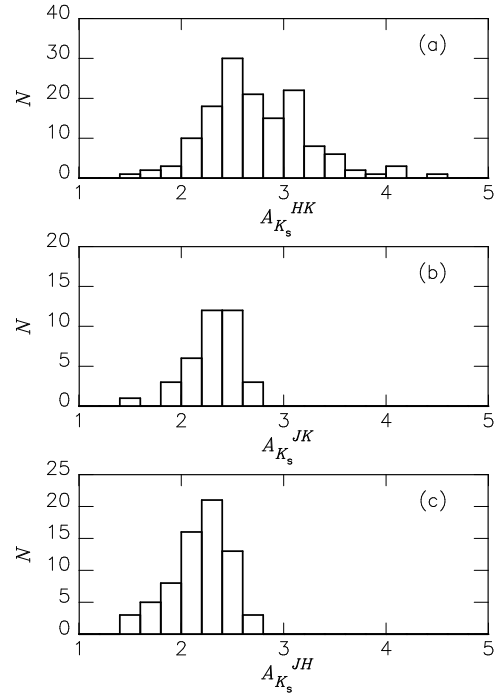}
\caption{
Histograms of $\AK$ values obtained by using
magnitudes in different pairs of filters.
\label{fig:AKhist}}
\end{center}
\end{minipage}
\end{center}
\end{figure}

\begin{figure}
\begin{center}
\begin{minipage}{75mm}
\begin{center}
\includegraphics[clip,width=0.85\hsize]{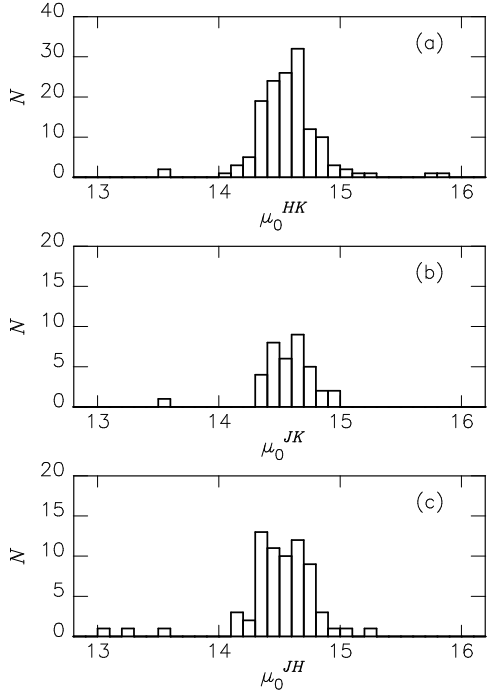}
\caption{
Same as Fig.~\ref{fig:AKhist}, but for $\mu_0$.
\label{fig:DMhist}}
\end{center}
\end{minipage}
\end{center}
\end{figure}

We have plotted histograms of $\AK$ in Fig.~\ref{fig:AKhist} and those of
$\mu_0$ in Fig.~\ref{fig:DMhist}, where each panel corresponds to the result
for each pair of filter bands. An obvious difference between the results
from different pairs is that Miras with $\AK>2.5$~mag are well detected in
panel (a) of Fig.~\ref{fig:AKhist} but not in panels (b) and (c). This is
because such highly-reddened Miras are too faint in $J$ for our survey. In
fact, none of the Miras with $\AK^{~HK}>3$~mag have $J$-band magnitudes.

\subsubsection{A note on small-amplitude variables\label{sec:SmallAmp}}

We did not take periodic variables with small-amplitudes
($\Delta J < 0.4$, $\Delta H < 0.4$, or $\Delta \Ks < 0.4$)
to be Miras. Among the sources with periods between 
100 and 350~d, 27 are of small-amplitude in $H$ and/or $\Ks$, i.e.\ 
$\Delta H < 0.4$ and/or $\Delta \Ks < 0.4$. If we estimate their distances
assuming that all of them are Miras, the resultant distribution of
$\mu_0^{~HK}$ looks different from that in the panel (a) of
Fig.~\ref{fig:DMhist}. In the case of Fig.~\ref{fig:DMhist}, there are almost no
objects at $\mu_0^{~HK}<14$.
In contrast,
five out of the 27
small-amplitude variables would be considered to be closer than
$\mu_0^{~HK}=14$~mag using (\ref{eq:PLRH}), (\ref{eq:PLRK}) and
(\ref{eq:solutionMHK}). Their distance moduli could be around 14.5~mag or
slightly larger than that if they follow the sequence B or C$^\prime$ in the
$\log P$-$\Ks$ diagram \citep{Ita-2004a} rather than the sequence C. There
is no reason to suppose that small-amplitude Miras are preferentially
located closer to us; thus it is likely that at least these five objects are
not Miras. We decided for safety to exclude the small-amplitude variables
from the list of Miras (even though many of them may follow the sequence C).
As it happens, our estimation of the distance to the GC is not affected even
if we include the small-amplitude variables in the following discussion.

\subsection{Errors in estimation of $\DMAK$}
\label{sec:ErrorSize}

As has been discussed in the preceding section, the solutions of
(\ref{eq:solutionAHK}) and (\ref{eq:solutionMHK}) are affected by the
uncertainties in our photometric error and the width of the PLR. We
calculated their effects on $\AK^{~HK}$ and $\mu_0^{~HK}$ and listed them in
Table~\ref{tab:DMAKerror}. We now summarize the sources of error in
estimating $\DMAK$, when both of the above are included, in
Table~\ref{tab:errors}. We classify the error sources into three types:
random, systematic and statistical. The first type relates to individual
Miras, while the second has a common effect on all the Miras in our sample.
When we study the distance to the GC, the statistical error carries more
weight than the random errors for individual objects because we make use of
many objects.

Our estimated periods have uncertainties of 0.025~dex in $\log P$ as seen in
section~\ref{sec:Pestimate}. They introduce uncertainties of 0.07--0.09~mag
in the predicted absolute magnitudes via
eqs.~(\ref{eq:PLRJ})--(\ref{eq:PLRK}). Since errors added to $M_J$, $M_H$,
and $\MK$ have the same sign, their effect on $\AK^{~HK}$ is almost
negligible ($\sim 0.01$~mag). On the other hand, $\mu_0^{~HK}$ is affected
by $\pm 0.09$~mag. This error is random and must be included with the
photometric error and that from the width of the PLR.
The result is a total error of
0.21~mag which randomly affects the distance estimates for individual Miras.

There are also uncertainties in the zero points of both the I04 catalogue
and the photometry of \citet{Nishiyama-2006a} which we used for the
calibration. The difference between the zero points in these works is
expected to be small because both were calibrated against standard stars in
\citet{Persson-1998}. The standard star \#9172 was used by
\citet{Nishiyama-2006a} and the uncertainty of their magnitude scale is
rather small (see section 2.1 in \citealt{Nishiyama-2009}). On the other
hand, I04 used a few dozen of Persson's standards. They used
$K$-band, not $\Ks$-band, magnitudes of these standard stars, but the
differences are negligible. The zero-point uncertainties were not given, but
are not expected to be larger than those in \citet{Nishiyama-2006a}. We
therefore adopted 0.01~mag for the uncertainties in the magnitude scale.

The extinction coefficients $r_J$ and $r_H$ give rise to uncertainties of
0.06~mag in $\AK^{~HK}$ and 0.07~mag in $\mu_0^{~HK}$ as already mentioned.
Here we do not consider any variation of extinction coefficient
with the line of sight in our survey region. Recently,
\citet{Gosling-2009} have suggested that there are
substantial variation on a very small scale ($\sim 5$~arcsec).
Such a variation would lead to different values
of $\AK$ from different pairs of filters. In contrast, the scatters
in Fig.~\ref{fig:DMAKcomp} can be explained without the variation.
This suggests that $r_J$ and $r_H$ are almost constant in our region.

We need at this point to consider the uncertainties arising from the
calibration of the PLR. We derived eqs.~(\ref{eq:PLRJ})--(\ref{eq:PLRK})
assuming the distance modulus to the LMC to be $\MuLMC=18.45$~mag. The
$\MuLMC$ value is one of the most fundamental values in the cosmic distance
scale and has been widely discussed. Most of the recent measurements are
around 18.4--18.5~mag (e.g.\ \citealt{vanLeeuwen-2007};
\citealt{Grocholski-2007}; Clement, Xu \& Muzzin, 2008). %%\citealt{Clement-2008}).
We adopt $\MuLMC$ and
its systematic error as being 18.45$\pm 0.05$~mag. A recent result based on
the PLR of type II Cepheids in the LMC is also consistent with this value
%% \citep{Matsunaga-2009}.
(Matsunaga, Feast \& Menzies, 2009).
Note that this error has nothing to do with the
estimation of $\AK$.

It may be asked whether there is a population effect on the PLR of Miras
that might also introduce a systematic error. \citet{Whitelock-2008} has
discussed the present status of the calibration of the PLR although they
focus on the $\log P$-$K$ relation and do not include other wavelengths.
They compared the PLR of the LMC Miras with those in the solar
neighbourhood, globular clusters and the Galactic bulge. The relations for
these environments agree with each other within an uncertainty of around
0.1~mag. Unfortunately, this uncertainty includes the error in $\MuLMC$
because one cannot make the above comparisons without its use. Their
conclusion is that population effects on the PLR are negligible
within the current accuracy.
As for the period-colour relation, some authors have reported systematic
differences between Miras in different stellar populations. For example,
\citet{Feast-1996} compared the $\log P$-$(J-K)_0$ relation of Miras in
different stellar systems and found a difference as large as 0.05~mag in
$(J-K)_0$ in the period range between 100 and 350~d. A shift of 0.05 in
$(J-K)_0$ would lead to a shift of 0.025~mag in the estimate of $\DMAK$.
%% Instrumental differences can have a large effect as we discussed in
%% section~\ref{sec:LMCPLR}.
We will assume here that the population difference between the LMC
and the Galactic bulge may introduce a colour difference of 0.05~mag in
$H-\Ks$ based on the top panel of Fig.~\ref{fig:compPCR} and that the zero
point of the PLR in $\Ks$ is not affected.
Thus uncertainties of 0.07~mag are introduced into both
$\AK^{~HK}$ and $\mu_0^{~HK}$.

In summary, there are four major sources of systematic error as listed in
Table \ref{tab:errors}, and their total size is expected to be 0.11~mag.

\begin{table}
\begin{minipage}{70mm}
\caption{Sources of error and their amounts for $\AK^{~HK}$ and $\mu_0^{~HK}$.
\label{tab:errors}}
\begin{center}
\begin{tabular}{lcc}
\hline
Error source & \multicolumn{2}{c}{Error size~(mag)} \\
 & $\AK^{~HK}$ & $\mu_0^{~HK}$ \\
\hline
\multicolumn{3}{c}{{\it Random uncertainty}} \\
photometric error               & 0.04 & 0.04  \\
width of the PLR                & 0.08 & 0.18  \\
period                          & 0.01 & 0.09  \\
\multicolumn{1}{r}{{\it total}} & 0.09 & 0.21  \\
\hline
\multicolumn{3}{c}{{\it Systematic uncertainty}} \\
photometric zero point          & 0.01 & 0.01  \\
extinction coefficient          & 0.06 & 0.07  \\
the LMC distance                & 0.00 & 0.05  \\
population effect on the PLR    & 0.07 & 0.07  \\
\multicolumn{1}{r}{{\it total}} & 0.09 & 0.11  \\
\hline
\multicolumn{3}{c}{{\it Statistical uncertainty}} \\
standard deviation of mean      &      & 0.02  \\
\multicolumn{1}{r}{{\it total}} &      & 0.02  \\
\hline
\end{tabular}
\end{center}
\end{minipage}
\end{table}

\subsection{Distance to the GC\label{sec:Distance}}

We have discussed the uncertainties in our distance estimation in the
preceding section. While random errors in $\mu_0$ were found to be about
0.2~mag, the peak at around $\mu_0=14.5$~mag has a standard deviation of
0.19~mag, which is obtained for the objects between 14 and 15~mag in
Fig.~\ref{fig:DMhist}. Therefore, the width of the distribution can be
explained by the random errors. This carries the implication that
the Miras are strongly concentrated to the GC.

Our survey field extends to approximately 12~arcmin in radius, corresponding
to 30~pc at the distance of the GC. This region is well within the so-called
Nuclear Bulge which is distinguishable from the large-scale Galactic bulge
\citep{Serabyn-1996}. The Nuclear Bulge consists of an $R^{-2}$ stellar
cluster which contains a significant stellar population as young as
$10^{7-8}$~yr. Long-period Miras (say $P > 600$~d)
and OH/IR stars indeed show a strong concentration towards the central
100~pc region %% (\citealt{Lindqvist-1992};
(Lindqvist, Habing \& Winnberg, 1992; %% \citealt{Wood-1998}).
Wood, Habing \& McGregor, 1998). However, it
is known that short-period Miras are relatively old (see the review by
\citealt{Feast-2009}), so that the Miras with $P=100$--350~d whose distances
we obtained may have a different spatial distribution from the younger
populations. %% \citet{Frogel-1999}
Frogel, Tiede \& Kuchinski (1999) showed that the younger component of the
stellar population is more concentrated than the overall
bulge population. Nevertheless, a large fraction of the old stellar
population is also located in a rather narrow region. As a rough estimate,
more than eighty percent of objects projected in our fields of view are
expected to be within 300~pc of the GC if we assume that their density
distribution follows the truncated power law used in
Binney, Gerhard \& Spergel (1997, see their eq.\ 1).
In contrast, the standard deviation around the peak in Fig.~\ref{fig:DMhist}
corresponds to $\pm 700$~pc. We therefore conclude that most of the Miras
that we have detected lie at the same distance and are centred around the GC
within our accuracy, so that the distance to the GC can be derived
legitimately from their distance statistics.

\begin{figure*}
\begin{minipage}{140mm}
\begin{center}
\includegraphics[clip,width=0.90\hsize]{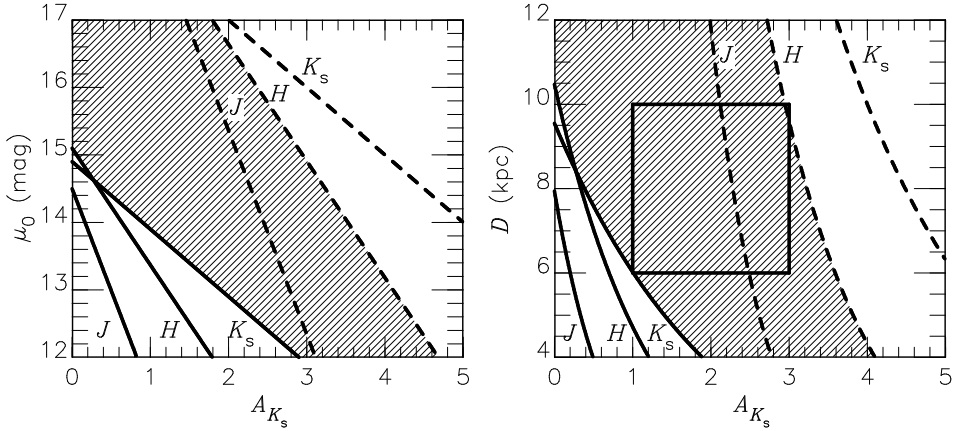}
\caption{
Schematic diagrams to show the parameter space of $\DMAK$
where we can detect Miras in $\JHK$. Dashed curves show
the detection limits while solid curves show the saturation limits
for $\JHK$ as indicated.
In the online version, blue, green and red curves corresponds to the limits
in $J$, $H$ and $\Ks$ respectively.
The two panels plot the same curves, but the $y$ axis in the right one is
on a linear scale. The rectangle in the right-hand panel indicates
the region we use to estimate the distance to the GC.
\label{fig:DMAKrange}}
\end{center}
\end{minipage}
\end{figure*}

It is necessary to consider any further biases which may have arisen. As
mentioned in section~\ref{sec:mean}, we could not always detect the Miras in
all three filters. The distances which we can estimate are limited
by the observable magnitude range and are also by the extinction
$\AK$. Based on the limits in $\JHK$ (Fig.~\ref{fig:Mhist}), it is possible
to define the range of $\DMAK$ over which one can observe objects with a
certain absolute magnitude. We here adopt $M_J=-5.0$, $M_H=-5.6$ and
$M_K=-5.9$ as typical magnitudes for Miras with $P=100$--$350$~d. In
Fig.~\ref{fig:DMAKrange}, we plot boundaries which correspond to the
detection limits (dashed curves) and the saturation limits (solid curves) in
the $\DMAK$ plane. The boundaries are different for different filters
as shown in Fig.~\ref{fig:DMAKrange}. The
right-hand panel presents the same curves as the left panel but with a linear
distance scale to show more clearly the range we are interested in. The
$J$-band detection limit implies that we cannot detect Miras if they are
more heavily extincted than 2--2.5~mag in $\AK$.
Note that it is possible to
detect Miras over a wider range of distance even in the $J$-band if the
reddening is smaller than $\AK=2$~mag. This is why the differences between
the distributions are not large in Fig.~\ref{fig:DMhist}. One can observe
the Miras with larger $\AK$ by using $H$ and $\Ks$, and so the number of
$(\mu_0^{HK}, \AK^{HK})$ estimates is significantly larger than when $J$ is
involved. We therefore use only $\mu_0^{~HK}$ and $\AK^{~HK}$ to discuss the
distribution of Miras in what follows. The shaded
areas in Fig.~\ref{fig:DMAKrange} indicates the range of $\DMAK$ where we
can detect Miras in both $H$ and $\Ks$. When we investigate the distance
distribution of Miras, we exclude those with $\AK>3$~mag because the Miras
on the far side of the Galactic bulge may not be detected at such
high extinctions. One should bear in mind that such a limitation
can introduce a selection bias in the values of $\DMAK$ that we found. The
rectangle in the right-hand panel indicates the range we made use of in
determining the distance to the GC.

\begin{figure}
\begin{center}
\begin{minipage}{75mm}
\begin{center}
\includegraphics[clip,width=0.85\hsize]{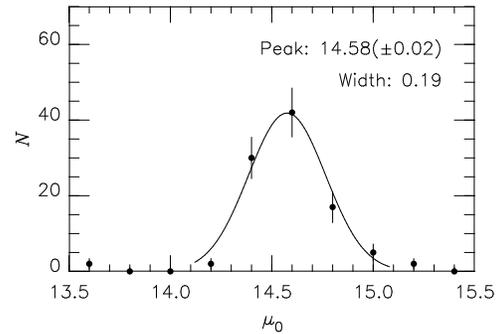}
\caption{
Distribution of the $\mu_0^{~HK}$ values obtained.
The error bar indicates the size of the Poisson noise ($\sqrt{N}$)
for each bin.
The mean value and width of the fitted Gaussian
are indicated in the panel.
\label{fig:fitDM}}
\end{center}
\end{minipage}
\end{center}
\end{figure}

We may now proceed to estimate the distance to the GC based on
the 100 Miras
which have estimated values of $\mu_0^{~HK}$ between 14 and 15~mag and
$\AK^{~HK}$ between 1 and 3~mag.  The latter condition is set to avoid
selection bias in the $\DMAK$ plane (Fig.~\ref{fig:DMAKrange}). One can
expect the spread in the estimated distance moduli to accord with the random
errors. As shown in Fig.~\ref{fig:fitDM}, the distribution is well fitted
by a Gaussian function with a peak of $\mu_0=14.58$~mag. The peak and the
width from the fitting are insensitive to the bin width. As already mentioned,
random errors for individual Miras are not important when we discuss the
distance to the GC because they average out. We take the standard deviation
of the mean, {\it viz.} 0.02~mag, as the uncertainty in the distance. As a
result, the distance modulus to the GC is estimated to be $14.58 \pm
0.02~{\it (stat.)}
\pm 0.11~{\it (syst.)}$~mag where the systematic error is included 
(section~\ref{sec:ErrorSize}). 
We denote the linear distance as $R_0$ and compare our
result, $R_0=8.24 \pm 0.08 \pm 0.42$~kpc, with those of others. When
necessary, we re-calculate $R_0$ by using $\MuLMC=18.45$ (if the distance
scale in a given work depends on this quantity).

In earlier works $R_0$ values have been obtained by various methods. That
involving stellar orbits around the central black hole Sgr~A$^*$ is a
potentially powerful tool, but there still remains a significant
uncertainty. Although \citet{Eisenhauer-2005} derived
$R_0 = 7.62~\pm~0.35$~kpc, 
\citet{Trippe-2008} has found that systematic errors neglected in the 2005
paper may change the result. \citet{Ghez-2008} obtained $R_0=8.0\pm 0.6$~kpc
using the same method but based on a different set of data. The same authors
obtained $R_0 = 8.07 \pm 0.32~({\it stat.}) \pm 0.13~({\it syst.})$~kpc as a
statistical parallax using the kinematic properties of several hundred
stars around the GC. Results from other methods include:
\citeauthor{Nishiyama-2006b} (\citeyear{Nishiyama-2006b}; $R_0 = 7.52$~kpc,
red clump stars),
%% \citeauthor{Groenewegen-2008} (\citeyear{Groenewegen-2008};
Groenewegen, Udalski \& Bono (2008;
$R_0 = 7.75$~kpc, type II Cepheids and RR Lyr variables).
Our result is similar to the recent results by \citet{Trippe-2008} and
\citet{Ghez-2008}.

A few other authors have estimated $R_0$ from Miras. Pioneering work was
done by \citet{Glass-1982}, who conducted near-IR photometry for a few dozen
of Miras towards the Baade windows (Sgr~I and NGC~6522).
\citet{Glass-1995} investigated the Miras towards the Sgr~I field
after further monitoring. Their result was $R_0=8.3 \pm 0.7$~kpc,
but they mentioned that their $R_0$ value will increase by 0.2~kpc if
the Sgr~I field is part of bar at $\sim 45^\circ$ to the line of sight.
\citet{Groenewegen-2005} used the Miras found in the OGLE-II survey fields
which cover various lines of sight between $\pm 11^\circ$ in $l$, mainly at
$b\sim -3.5^\circ$. They found that the difference between the distance
moduli of the LMC and the Galactic bulge is 3.72~mag, which corresponds to
$R_0=8.9 \pm 0.7$~kpc. They also argued that $R_0$ can be as small as
8.35~kpc if one assumes a strong metal dependency of the PLR of Miras.
\citet{Glass-2009} adopted $R_0=8.5$~kpc by comparing their plots of mid-IR
magnitudes against periods for pulsating red giants in the LMC and the
Galactic bulge including semi-regulars on other sequences. The previous
results show rough agreements with ours considering the systematic
uncertainties. Note that the fields they examined are at least a few degrees
from the GC. Our report is the first to estimate $R_0$ by using the Miras
towards the GC region itself. Furthermore, the method described here has
made it possible to overcome the complicated effects of the interstellar
extinction.

\subsection{Interstellar extinction\label{sec:Extinction}}

\begin{figure*}
\begin{minipage}{150mm}
\begin{center}
\includegraphics[clip,width=0.95\hsize]{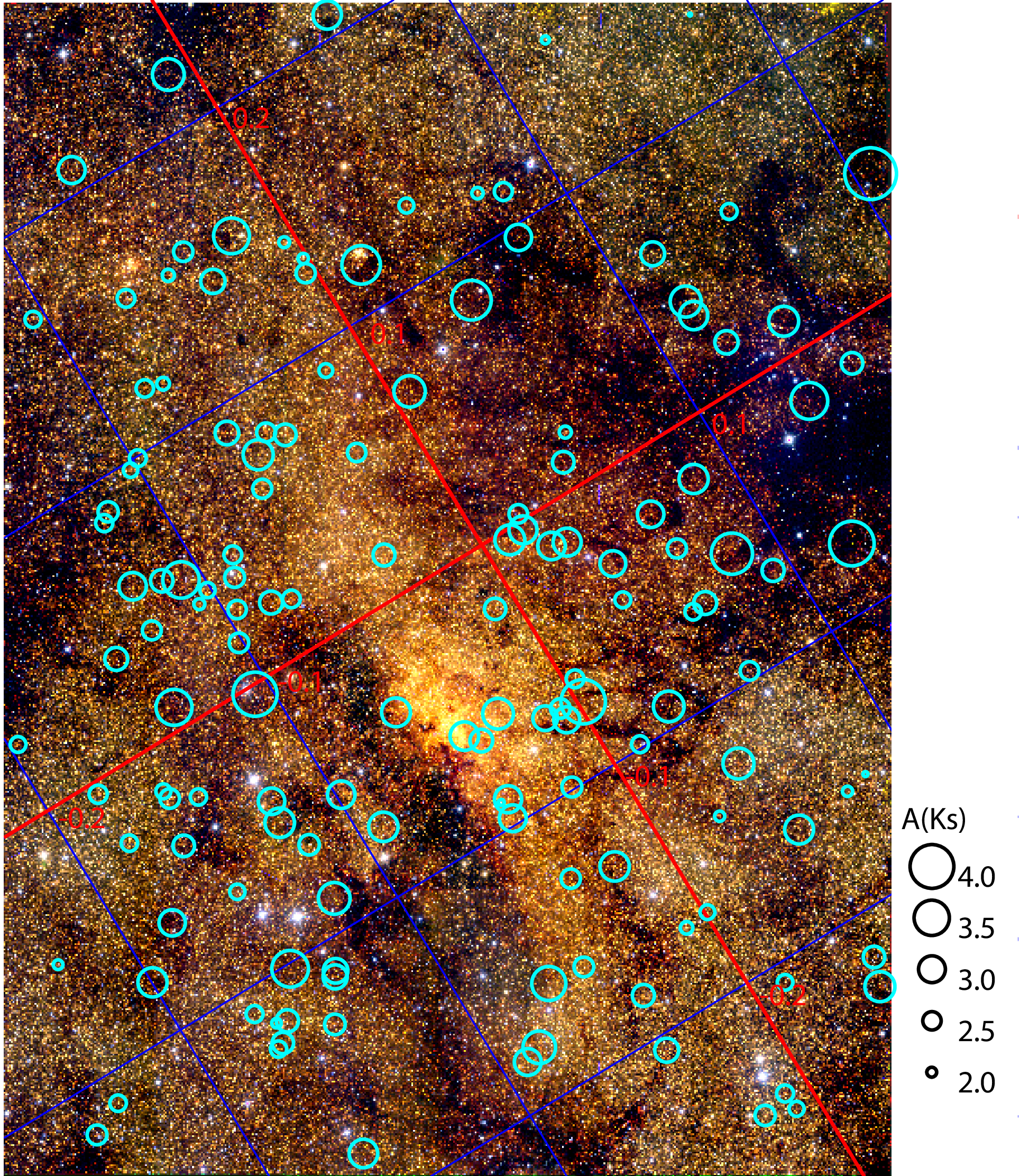}
\end{center}
\caption{
Locations of 143 Miras with $\AK^{~HK}$ obtained.
The obtained extinction for each object is indicated by the size of a circle
as shown besides the panel.
\label{fig:AKmap}}
\end{minipage}
\end{figure*}

In the course of this work, besides their distances, interstellar
extinctions have been estimated for more than 150 Miras. Because we estimate
both at the same time, this method has the potential to investigate the
extinction in three dimensions directly, revealing how it changes with
distance and with line of sight. In spite of the fact that our objects are
located at the same distance, the $\AK$ values cover a large range;
the foreground extinction varies strongly with the line of
sight. In Fig.~\ref{fig:AKmap}, we plot the locations of the Miras with
$\AK^{~HK}$ values in galactic coordinates. The sizes of the symbols vary
according to $\AK$ as indicated. Whilst circles with various sizes
scatter across our observed area, there are some clumps of
similar-sized circles.

\begin{figure}
\begin{center}
\begin{minipage}{75mm}
\begin{center}
\includegraphics[clip,width=0.85\hsize]{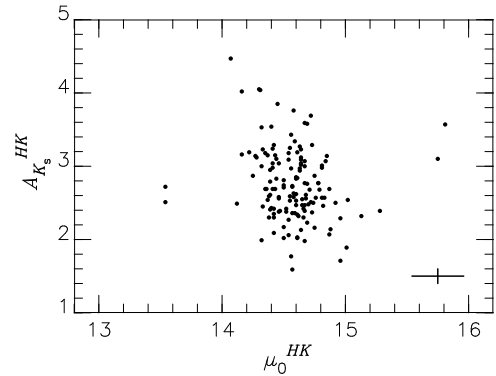}
\end{center}
\caption{
$\AK^{~HK}$ are plotted against $\mu_0^{~HK}$.
\label{fig:DMAK}}
\end{minipage}
\end{center}
\end{figure}

Fig.~\ref{fig:DMAK} plots $\AK^{~HK}$ against $\mu_0^{~HK}$. The cross
indicates the sizes of the errors in the two quantities. Although most of
the samples in this study are located at the same distance within the
accuracy, there are two sources closer and two further than the main
group. The $\AK$ values obtained for these objects
do not show systematic differences from
those for the objects with $\mu_0=14$--$15$~mag. Unfortunately, the number
of tracers with $\DMAK$ is too small to carry out a detailed analysis of the
three-dimensional extinction map. Also note that our sample is limited in
the $\DMAK$ plane (see Fig.~\ref{fig:DMAKrange}). Those with smaller
extinction and closer to the Sun will be saturated in our survey, while
those with larger extinction and further than the GC are expected to be too
faint.

\begin{figure}
\begin{center}
\begin{minipage}{75mm}
\begin{center}
\includegraphics[clip,width=0.75\hsize]{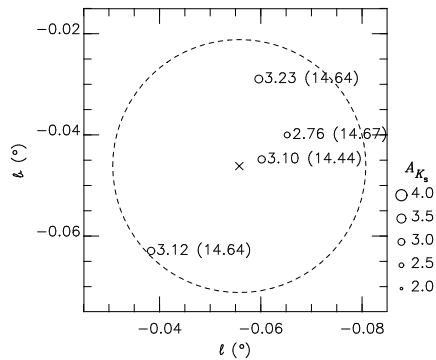}
\end{center}
\caption{
This is similar to Fig.~\ref{fig:AKmap},
but is a closeup for the proximity of Sgr~A$^*$
which is indicated by the cross.
The dashed circle has a radius of 1.5~arcmin and a centre at Sgr~A$^*$.
The $\AK$ (and $\mu_0$ in parentheses) is indicated for each object.
\label{fig:AKmapGC}}
\end{minipage}
\end{center}
\end{figure}

We here estimate the foreground interstellar extinction towards Sgr~A$^*$
($l = -0.05575^\circ, b = -0.04617^\circ$). Fig.~\ref{fig:AKmapGC} shows the
locations and the $\AK^{~HK}$ values (and $\mu_0^{~HK}$ in parentheses)
for the Miras
around Sgr~A$^*$. Four Miras are located within a radius of 1.5~arcmin,
indicated by the dashed circle. It is suggested that
the $\AK$ in the direction towards Sgr~A$^*$ is approximately 3~mag.
Our value corresponds to $A_V \sim
45$~mag if we use the extinction coefficient of $A_V/\AK=16$ obtained by
\citet{Nishiyama-2008}. In contrast, extinction values of 2.6 and
2.8~mag in $\AK$ were obtained toward Sgr~A$^*$ by \citet{Eisenhauer-2005}
and \citet{Schodel-2007} with higher angular resolution. The latter found
that the extinction changes on rather small scales. Our estimate should
therefore be regarded as a typical value for a larger region surrounding
Sgr~A$^*$ rather than the value in its precise direction.

\subsection{Mid-IR PLR}

\citet{Glass-2009} recently investigated the PLR of Miras in the mid-IR
range. Their samples consist of the Miras towards the LMC and the NGC~6522
Baade window. \citet{Schultheis-2009}
fitted the mid-IR PLRs for the G01 Miras around the GC,
but there remain large scatters because it was not
possible to correct for the variable interstellar extinction. 

\begin{figure}
\begin{center}
\begin{minipage}{75mm}
\begin{center}
\includegraphics[clip,width=0.85\hsize]{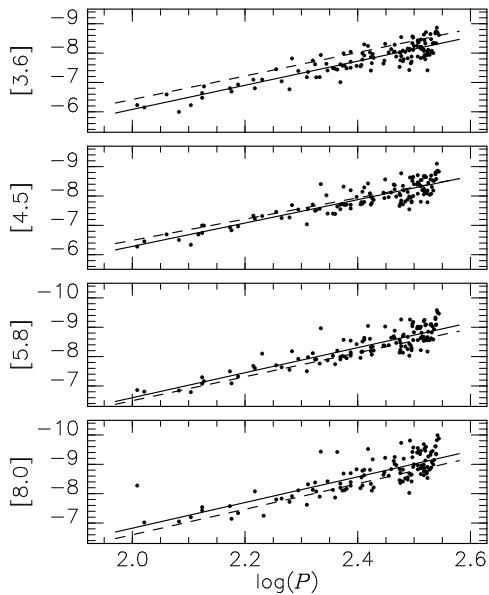}
\end{center}
\caption{
Period-luminosity relation in {\it Spitzer}/IRAC filters.
The fitted relations are shown as solid lines,
while the PLR in \citet{Glass-2009} are indicated by dashed lines.
Note that they adopted a distance differing from ours by 0.12~mag
and we have shifted their fits by this amount.
\label{fig:IRACPLR}}
\end{minipage}
\end{center}
\end{figure}

We have plotted the IRAC magnitudes from \citet{Ramirez-2008}
against the periods
in Fig.~\ref{fig:IRACPLR} for our Miras with $\DMAK$. The
fitted linear relations are:
\begin{eqnarray}
\IRACa = -4.14 (\pm 0.20) (\log P -2.3) - 7.32 (\pm 0.03),
 \label{eq:PLR36} \\
\IRACb = -4.03 (\pm 0.20) (\log P -2.3) - 7.48 (\pm 0.03),
 \label{eq:PLR45} \\
\IRACc = -4.33 (\pm 0.24) (\log P -2.3) - 7.88 (\pm 0.04),
 \label{eq:PLR58} \\
\IRACd = -4.46 (\pm 0.33) (\log P -2.3) - 8.14 (\pm 0.06),
 \label{eq:PLR80} 
\end{eqnarray}
indicated by the solid lines. The residual standard deviations around
(\ref{eq:PLR36}), (\ref{eq:PLR45}), (\ref{eq:PLR58}) and (\ref{eq:PLR80})
are 0.29, 0.29, 0.32 and 0.42~mag, respectively. We here used the
$\mu_0^{~HK}$ and $\AK^{~HK}$ obtained for individual Miras and the
extinction coefficients obtained by \citet{Nishiyama-2009} to get their
absolute magnitudes. Because most of the Miras are found at the same
distance within our accuracy, it is also reasonable to use the common
distance modulus of 14.58~mag.  Even if we do so, the resultant relations do not
change from eqs.~(\ref{eq:PLR36})--(\ref{eq:PLR80}) including their scatters.

The scatter around the PLRs that we have found are rather small considering
the uncertainty in the $\DMAK$ and the fact that the IRAC
magnitudes are based on a single-epoch survey. Uncertainties in the apparent
distance moduli in the IRAC filters are estimated at 0.18~mag for individual
Miras. The mid-IR amplitudes of Miras are not well known, but those
at around 3.5~$\mu\mathrm{m}$ can be as large as in the $K$-band
\citep{Feast-1982}. These uncertainties suggest that
the intrinsic widths of the
mid-IR PLRs are narrower than in Fig.~\ref{fig:IRACPLR}.

In Fig.~\ref{fig:IRACPLR}, we also plot the relation obtained for the
C-sequence by \citet{Glass-2009}
as the dashed lines.
Their adopted distance, 14.7~mag for the
NGC~6522 field, differs from ours by 0.12~mag and was compensated for
before plotting. The slopes agree well with ours. In contrast, the zero
points are slightly shifted depending on the wavelength; approximately by
$+0.3$~mag for the $\log P$-$[3.6]$ relation and $-0.2$~mag for the $\log
P$-$[8.0]$ relation. The reason for the different zero points is unclear.
Note that \citet{Glass-2009} included semi-regulars and long-period Miras
which may have mid-IR excesses in their sample.

\subsection{Future prospects}

As has been shown, it is possible to obtain both distances and extinctions
for Miras with $P=100$--$350$~d by making use of the PLR in two or more
near-IR filter bands. This is a very promising tool for investigating
Galactic structure. One of the major sources of error is the residual
scatter around the PLR. Because we do not have {\it a priori} information on
how a certain Mira deviates from the fitted PLR, the scatter introduces an
uncertainty in the distance estimation. We expect that the intrinsic
scatter is smaller than what we found in Fig.~\ref{fig:LMCPLR}. The
magnitudes used in Fig.~\ref{fig:LMCPLR} are intensity means of ten-epoch
monitoring data, which can introduce an uncertainty as large as 0.1~mag
based on the Monte-Carlo simulation presented in section~\ref{sec:mean}. The
scatter should get smaller if we take mean magnitudes more accurately.

The uncertainty in the zero point of the PLR and the possible population
effect may also introduce significant errors; it is important to note
however that these will be systematic ones. Future parallax observations of
Miras in the solar neighbourhood may reduce these uncertainties. For
example, \citet{Vlemmings-2003} and \citet{Vlemmings-2007} have reported the
parallaxes for five Miras with OH maser emission. As reviewed by
\citet{Whitelock-2008}, their results are consistent with the PLR of the LMC
Miras within their uncertainties, which are still not small enough to draw
strong conclusions about the zero point or the population effect. VERA, a
Japanese VLBI project, is one of the most promising projects for calibrating
the PLR of Miras in the Galaxy (Nakagawa~{et~al.} \citeyear{Nakagawa-2008},
\citeyear{Nakagawa-2009}). However, the use of radio parallaxes is
constrained by the small numbers of Miras with maser emission in the
short-period range, say $P \leq 300$~d. In our method for deriving the
distance and the extinction, we make use of short-period Miras to avoid the
effects of circumstellar extinction and/or hot-bottom-burning. In this
regard, future optical and near-IR astrometric satellites such as GAIA
\citep{Turon-2008} and JASMINE \citep{Gouda-2008} can play a fundamental
role in the calibration of the PLR.

\section{Summary\label{Summary}}

This paper reports 1364 variable stars in a $20\arcmin\times 30\arcmin$
region around the GC found through our near-IR survey covering seven years.
We have obtained periods for 549 objects, most of which are considered to be
Miras. We have described a method for obtaining distances and extinctions
for Miras with periods between 100 and 350~d by making use of their PLR. The
technique depends on knowing mean magnitudes in two or more filters in the
near IR. A particular effort has been made to assess the uncertainties; the
random error is about 0.2~mag in the distance modulus for an individual Mira,
while the systematic error is about 0.11~mag, mainly originating from
a combination of uncertainties in the LMC distance ($\pm 0.05$~mag), the
extinction coefficient ($\pm 0.07$~mag) and the population effect on the
PLR ($\pm 0.07$~mag). 

We obtained the distances of 143
Miras by using the $H$- and $\Ks$-band mean
magnitudes. They are strongly concentrated to the GC, whose distance modulus
was estimated at 14.58~mag. The statistical error is 0.02~mag from the
standard deviation of the mean and the systematic error is again 0.11~mag.
Our statistical error is the smallest among estimates to date, and the
distance we have obtained shows rough agreement with recent results from
other methods, within the systematic uncertainties. The PLR of Miras is
shown to be a powerful tool for studying Galactic structure, although there
remains a significant systematic error. It is expected that better
calibration of the PLR will improve the accuracy of the method; this may be
achieved by parallax projects such as VERA and GAIA.

We also have discussed the distribution of the interstellar extinction
towards the GC. Because we can estimate distances and
extinctions for individual Miras, they can be used as tracers to
reveal three-dimensional structure. It is possible to study how the
extinction changes with distance along various lines of sight, although the
low space density of Miras may prevent us from investigating small
structures. We have confirmed that the extinction towards the GC region
varies complicatedly between 1.5 and 4~mag in $\AK$, excepting towards
extremely dark nebulae, and we have estimated the extinction towards the
Sgr~A$^*$ region to be $\AK \sim 3$~mag.

\section*{Acknowledgments}

The authors acknowledges valuable comments of Michael Feast.
We thank the IRSF/SIRIUS team members for our near-IR observations. The
IRSF/SIRIUS project was initiated and supported by Nagoya University,
National Astronomical Observatory of Japan and University of Tokyo in
collaboration with South African Astronomical Observatory under a financial
support of Grant-in-Aid for Scientific Research on Priority Area (A) Nos.
10147207, 10147214, 15071204, 15340061, 19204018 and 21540240
of the Ministry of Education, Culture, Sports, Science
and Technology of Japan.

\bibliographystyle{mn2e}

%% \appendix

\label{lastpage}

\end{document}